# Interface control of the magnetic chirality in CoFeB|MgO heterosctructures with heavy metal underlayers

Jacob Torrejon[1], Junyeon Kim[1], Jaivardhan Sinha[1], Seiji Mitani[1] and Masamitsu Hayashi[1*]

[1]*National Institute for Materials Science, Tsukuba 305-0047, Japan*

Michihiko Yamanouchi[2,3] and Hideo Ohno[2,3,4]

[2] *Center for Spintronics Integrated Systems, Tohoku University, Sendai 980-8577, Japan*

[3]*Research Institute of Electrical Communication, Tohoku University, Sendai 980-8577, Japan*

[4]*WPI Advanced Institute for Materials Research, Tohoku University, Sendai 980-8577, Japan*

**Recent advances in the understanding of spin orbital effects in ultrathin magnetic heterostructures have opened new paradigms to control magnetic moments electrically[1,2]. The Dzyaloshinskii-Moriya interaction[3,4] (DMI) is said to play a key role in forming a Neel-type domain wall that can be driven by the spin Hall torque[5-8], a torque resulting from the spin current generated in a neighboring non-magnetic layer via the spin Hall effect[9,10]. Here we show that the strength and sign of the DMI can be changed by modifying the adjacent heavy metal underlayer (X) in perpendicularly magnetized X|CoFeB|MgO heterostructures. Albeit the same spin Hall angle, a domain wall moves along or against the electron flow depending on the underlayer. We find that the sense of rotation of a domain wall spiral[11] is reversed when the underlayer is changed from Hf to W and the strength of DMI varies as the number of 5d electrons of the heavy metal layer changes. The DMI can even be tuned by adding nitrogen to the underlayer, thus allowing interface engineering of the magnetic texture in ultrathin magnetic heterostructures.**

*Email: hayashi.masamitsu@nims.go.jp

Understanding the underlying physics of current driven domain wall motion is essential in developing advanced storage class memory devices[12]. Conventionally, domain walls move along the electron flow (against the current) when driven by spin transfer torque[13,14]. Recently, a number of experiments have shown that the domain walls can instead move against the electron flow in magnetic heterostructures[7,8,15-19]. To describe this effect, the spin Hall effect[9,10] in the heavy metal layer has been considered as a possible source of the spin current. The generated spin current diffuses into the ultrathin magnetic layer and exerts torque, termed the "spin Hall torque[2]", on the domain wall magnetization only if the wall forms a Neel type wall[5,6]. To move sequences of domain walls with current in the same direction, the Neel wall has to alternate its chirality between neighboring domain walls. This requires formation of a "domain wall spiral[11]", which can be generated in systems with large spin orbit coupling and broken inversion symmetry via the Dzyaloshinskii-Moriya interaction[3,4] (DMI).

In the above model[5-8], the direction to which a domain wall moves with current is determined by the signs of the spin Hall effect and the DMI. The sign of the spin Hall effect depends on the heavy metal layer and is determined by the element specific spin orbit coupling; for example, it is opposite[2,8,20,21] for Pt and Ta. For the DMI, the sign depends upon the spin orbit coupling as well as the structural symmetry of the magnetic layer[3,4]. For example, in three dimensional bulk-like systems, the sense of rotation of the magnetic structure, i.e. the "chirality", can either follow or be opposite to the crystallographic chirality in Mn and Fe based non-centrosymmetric B20 structures[22,23], respectively, indicating the difference in the spin orbit coupling of the Mn and Fe based systems. The magnetic chirality at surfaces has been studied intensively using spin polarized scanning tunneling microscopy[24,25]. Here the surface atomic configuration plays an important role in setting the chirality.

The origin of the DMI at interfaces is more difficult to treat as the structural symmetry determination is non-trivial. It has been reported that DMI changes its sign depending on the order of the film stack[7,26], which is consistent with the three-site indirect exchange mechanism[27,28] proposed previously. Recent experiments[8] have indicated that for a given magnetic layer (CoFe) the sign of the DMI is the same even when the adjacent non-magnetic layer (Pt or Ta) has opposite sign of the spin orbit coupling constant.

Here we show that the size and sign of the DMI can be changed for a given magnetic layer when the neighboring non-magnetic layer is modified. In X|CoFeB|MgO heterostructures with different heavy metal underlayers (X), we find that the domain wall moves along or against the electron flow depending on the underlayer material. The sign of the spin Hall angle is the same for all underlayers, indicating that the sign of bulk spin orbit coupling constant of X is the same. In contrast, the magnetic chirality of the domain walls is reversed when the underlayer is changed from Hf to W. The strength of the DMI varies as the number of $5d$ electrons the neighboring layer (X) carries is changed, and it can be tuned by, for example, adding nitrogen to Ta which can possibly influence the number of $5d$ electrons by orbital hybridization (nitrogen $2p$ and Ta $5d$).

Films are deposited using magnetron sputtering on Si(001) substrates coated with 100 nm thick thermally oxidized Si. The film stack is Sub.|$d$ X|1 CoFeB|2 MgO|1 Ta (units in nanometer). We use Hf, Ta, TaN and W for the underlayer material X. TaN is formed by reactively sputtering Ta in a mixture of $N_2$ and Ar gas[29]. $Q$ represents the fraction of $N_2$ gas introduced during the Ta sputtering with respect to the entire ($N_2$+Ar) gas. The composition of TaN is close to ~50 at% N for all $Q$ explored (up to ~2.5 %). The thickness ($d$) of the underlayer X is varied within in the substrate using a linear shutter during the sputtering. Magnetic easy axis points out of plane owing to the perpendicular magnetic anisotropy developed at the CoFeB|MgO interface[30].

Wires are patterned using optical lithography and Ar ion etching followed by a lift-off process to form the electrical contacts made of 10 Ta|100 Au. An optical microscopy image of the wire along with schematic illustration of the experimental setup is shown in Fig. 1(a) and 1(b). Variable amplitude voltage pulses (duration fixed to 100 ns) are fed into the wire from a pulse generator. Positive voltage pulse supplies current into the wire that flows along the +x direction. A domain wall is nucleated[31] by applying a voltage pulse above a critical amplitude which depends on the film stack. Kerr microscopy is used to acquire magnetic images of the sample. Current driven domain wall velocity is estimated by dividing the distance the wall traveled, obtained from the Kerr images, by the total pulse length. Typical wires studied here have a width of ~5 μm. Representative devices are measured for studying current induced domain wall motion.

Exemplary Kerr microscopy images are shown in Fig. 1(c) and 1(d) when negative and positive voltage pulses are applied to devices made of Ta and TaN(Q=0.7%) underlayers, respectively. The domain wall moves along the electron flow for the former whereas it moves against the electron flow for the latter. Note that the domain walls shown in Fig. 1(c)-(d) correspond to same domain configuration (↓↑ walls). Depending on the thickness and dielectric constant of the each layer including the 100 nm thick $SiO_2$, the Kerr contrast can change (see supplementary information S2.A).

Domain wall velocity as a function of the voltage pulse amplitude is summarized in Fig. 2. Positive velocity corresponds to a domain wall moving toward the +x direction. For Hf and Ta underlayer films, the domain wall always moves along the electron flow. This also applies for the thin TaN underlayer films. However, the domain wall moves against the electron flow for the thicker TaN underlayer films and for all of the W underlayer films. Note that the applicable pulse amplitude is limited by the voltage at which current induced domain wall nucleation occurs.

The threshold current density needed to move a domain wall is plotted in Fig. 3(a-h), open symbols. The current density that flows in the CoFeB layer ($J_C^{CoFeB}$) and the underlayer ($J_C^{Under}$) are shown in the top (a-d) and bottom (e-h) panels, respectively. The solid symbols in Fig. 3 represent the *maximum* current density applied to each device; beyond this current density, we find evidence of current induced domain nucleation. Thus the current density applied to each device range between the open and solid symbols. In almost all cases, the threshold current density (both $J_C^{CoFeB}$ and $J_C^{Under}$) decreases as $d$ is increased.

The out of plane field needed to move a domain wall, i.e. the propagation field ($H_P$), is plotted in Fig. 3(i-l). $H_P$ represents the strength of domain wall pinning along the wire. The change in $H_P$ is mostly correlated with the magnetic anisotropy of the films: films with larger $K_{EFF}$ display larger $H_P$. The origin of the linear $d$ dependence of $H_P$ for the TaN films is not well understood. $H_P$ is generally small and is below ~30 Oe for all devices.

To examine the underlying mechanism of current driven domain wall motion, we study the size and sign of spin Hall torque (i.e. the current induced effective magnetic field) in a Hall bar patterned on the same substrate using the adiabatic (low frequency) harmonic Hall voltage measurements[32-34]. Figure 4 shows the effective field components directed transverse to ($\Delta H_T/J^{Under}$, Fig. 4(a-d)) and along ($\Delta H_L/J^{Under}$, Fig. 4(e-h)) the current flow direction plotted as a function of the underlayer thickness $d$. $\Delta H_{T(L)}/J^{Under}$ represents the effective field normalized by the current density ($1 \times 10^8$ A/cm$^2$) flowing through the underlayer. Measurements are performed with current density smaller than ~$2 \times 10^7$ A/cm$^2$. The $d$ dependence of the longitudinal effective field is similar among the film structures studied here: $\Delta H_L/J^{Under}$ increases in magnitude with increasing $d$ and saturates at a certain $d$ (the origin of the drop at large $d$ for the Hf underlayer films is not clear, see supplementary information S2.D for discussion). The source of the effective field is thus likely the spin Hall torque[2], whose direction is determined by the spin Hall angle. Since the sign of $\Delta H_L/J^{Under}$ at saturation is the same for all underlayer films, these results show that the sign of the spin Hall angle is the same. Note that $\Delta H_L/J^{Under}$ for large $d$ of the W underlayer films is nearly ~8 times larger than that of the Ta and TaN underlayer films.

The transverse component ($\Delta H_T/J^{Under}$) also increases with $d$ for the Hf, Ta and TaN underlayer films. However, the $d$ dependence of $\Delta H_T/J^{Under}$ for the W underlayer films is different from the others: $\Delta H_T/J^{Under}$ changes its direction at $d$~2.5 nm when $\Delta H_L/J^{Under}$ is more or less constant. This is in stark contrast to the Hf, Ta and TaN underlayer films in which $\Delta H_T/J^{Under}$ and $\Delta H_L/J^{Under}$ change their direction at similar $d$ (and typically at small $d$)[33]. Understanding the origin of $\Delta H_T/J^{Under}$ in all underlayer films, in particular for W, requires further (theoretical) study. With regard to current induced domain wall motion, it is predominantly the longitudinal component that drives a Neel wall[5,6].

The spin Hall angle $\theta_{SH}$ and the spin diffusion length $\lambda_{SD}$ estimated from $\Delta H_L/J^{Under}$ are shown in Fig. 6(a) and 6(b), respectively, as a function of the underlayer material. Note that we assume a transparent interface between the underlayer and the CoFeB layer for spin transmission to simplify the procedure[2] of estimating $\theta_{SH}$. However, the presence of large $\Delta H_T/J^{Under}$ suggests that the interface is not transparent. Thus the $\theta_{SH}$ presented here is to provide a rough estimate. We find a large $\theta_{SH}$ for the W underlayer films, similar to what has been reported recently[35]. $\theta_{SH}$ for the Ta underlayer film lies between numbers reported using different film structures and different evaluation methods[2,20] (see discussions of $\theta_{SH}$ in the supplementary material S2.D). The change in $\theta_{SH}$ with the $N_2$ gas concentration ($Q$) for the TaN underlayer films may be partly related to the boron concentration in TaN which decreases with increasing nitrogen concentration[29] and can influence the scattering rate (and thus the spin diffusion length). The spin diffusion length roughly scales with the resistivity of the underlayer film but also is dependent on each element.

As the sign of the spin Hall angle is the same for all film structures, we infer that the DMI is changing its sign between different underlayers. In out of plane magnetized systems, the preferred domain wall configuration is the Bloch type for the wire dimension used here: a Neel wall is only stable for narrow wires (typically below ~100 nm) where shape anisotropy starts to dominate[14,36]. However, the DMI can promote a Neel type wall with a fixed chirality. This interaction can be modeled as an additional offset field ($H_{DM}$) applied along the wire's long axis for a given domain wall[5,7,8]. The offset field changes its direction depending on whether the magnetization of the neighboring domain points ↑↓ or ↓↑, thus forming a domain wall spiral[11]. We thereby study the wall velocity as a function of the in-plane magnetic field directed along the wire's long axis ($H_L$) to probe $H_{DM}$[7,8].

Figure 5(a,b) show representative results of the wall velocity vs. $H_L$ for two different devices in which the domain wall moves in opposite directions when driven by current. The velocity scales almost linearly with $H_L$ in all cases. At zero $H_L$, both ↑↓ or ↓↑ walls move in the same direction for a given film structure. However, the field at which the velocity becomes zero (defined as $H_L^*$ hereafter) is different depending on the domain configuration (↑↓ or ↓↑ walls). For example, $H_L^*$ is positive (negative) for a ↑↓ (↓↑) wall when the wall moves along the electron flow (Fig. 5a). This indicates that there is a negative (positive) offset field ($H_{DM}$) associated with the ↑↓ (↓↑) wall. The direction of this offset field reverses when the wall moves against the electron flow (Fig. 5b). These results show that the domain wall spiral possess a left handed chirality (↑←↓ or ↓→↑) for the walls moving along the electron flow and it forms a right handed chirality (↑→↓ or ↓←↑) when the direction of the wall motion reverses.

The underlayer thickness dependence of $H_L^*$ is plotted in Fig. 5(c-f). We find a clear correlation between the direction of the wall motion (background color coding) and the sign of $H_L^*$. However, note that $H_L^*$ does not represent the DMI offset field ($H_{DM}$) when spin transfer torque is present. According to the one-dimensional (1D) model of a domain wall[5,7,8,37], $H_L^*$ for a ↑↓ wall is expressed as (see supplementary material S3.B):

$$H_L^* = -\left[D + \text{sgn}(\theta_{SH})0.21P\left|J^{CoFeB}\right|\right]\frac{1}{M_S\Delta} \qquad (1)$$

where $D(=H_{DM}M_S\Delta)$ is the Dzyaloshinskii-Moriya (anisotropic) exchange constant, $M_S$ is the saturation magnetization, $\Delta = \sqrt{A/K_{EFF}}$ is the domain wall width parameter and $P$ is the spin polarization of the current that flows in the CoFeB layer. We use the exchange constant ($A$) of the CoFeB layer obtained in a similar heterostructure[38] and the variation of the magnetic anisotropy ($K_{EFF}$) with $d$, shown in Fig. S1 of the supplementary material, is taken into account in estimating $\Delta$. The unit of $J^{CoFeB}$ is $10^8$ A/cm$^2$. The non-adiabatic spin torque contribution in the CoFeB layer is assumed to be negligible, as reported previously in a similar system[39]. We fit $H_L^*$ vs. $d$ using Eq. (1) to estimate $D$. Results are shown by the solid/dashed lines in Fig. 5(c-f). Note that when the spin Hall angle is zero, $H_L^*$ is not well defined. Thus films with small $\Delta H_L$ (smaller than ~10 Oe in Fig. 4(e-h)) are not considered in the fitting. The change in the sign of $H_L^*$ for the thin TaN underlayer films can be explained well by considering contribution from the spin transfer torque.

The underlayer dependence of $D$ is plotted in Fig. 6(c). The symbols show $D$ when $P$ is set to 0.7. The error bars indicate variation of $D$ when spin transfer torque is absent (lower bound of the error bars) or when $P$ is 1 (upper bound) for the Ta and TaN underlayer films. For the Hf and W underlayer films, the error bars indicate the variation of $D$ with the fitting since contribution from spin transfer torque is relatively small. As evident, $D$ varies as one moves along the 5th row of the periodic table from Hf to W: Hf underlayer films possess a negative $D$ (left handed magnetic chirality), whereas $D$ is positive for TaN and W (right handed). D is small for Ta underlayer films. It should be noted that when a large domain wall pinning is present (i.e. large $H_P$), $H_L^*$ measured using a linear fitting of the velocity vs. $H_L$ (Fig. 5(a,b)) can underestimate the offset field $H_{DM}$ (see Fig. S8 and section S3.C in the supplementary material). Thus the values provided in Fig. 6(c) correspond to the lower limit of $|D|$.

The change in the size and sign of the DMI with the heavy metal elements may be related to the change in the charge localization of the interface atoms, which has been reported to change the sign of the Rashba spin splitting at metal alloy surfaces[40]. It is somewhat surprising that addition of nitrogen to Ta can influence the DMI in such a dramatic way. First principle calculations show[41,42] that the valence (5$d$-like) electron density vary due to strong hybridization of the N $p$ and Ta $d$ orbitals in transition metal nitrides. Thus TaN may carry more 5$d$-like electrons than Ta, thus contributing to the change in DMI. There is also a possibility that the atomic configuration, e.g. local atomic arrangement and/or the amount of boron present at the interface[29] or, varies as we change the underlayer material, and thus contribute to the change in $D$.

We finally describe the threshold current needed to move domain walls with spin Hall torque. Since the effect of spin Hall torque is similar to that of an out of plane magnetic field[5], the spin Hall effective field ($\Delta H_L$) only needs to overcome the propagation field ($H_P$) to trigger domain wall motion provided that the wall configuration is a Neel-type, i.e. the threshold longitudinal effective field $\Delta H_L^C$ required to move a wall should equal $H_P$. Experimentally $\Delta H_L^C$ can be calculated using the data from Fig. 3(e-h) and Fig. 4(e-h), i.e. $\Delta H_L^C = \Delta H_L / J^{Under} \times J_C^{Under}$. We study $\Delta H_L^C$ for underlayer films with non-negligible $\Delta H_L$.

The symbols in Figure 6(d) show $\Delta H_L^C$ as a function of $H_P$. The solid line indicates the relationship between $\Delta H_L^C$ and $H_P$ for a Neel wall, i.e. $\Delta H_L^C=H_P$. Deviation from this relationship indicates that the wall is not a Neel wall, i.e. $H_{DM}$ is not large enough to overcome the demagnetization field and force a Neel wall (see supplementary material S3.D). We find that for the Hf and most of the TaN underlayer films, $\Delta H_L^C$ scales with $H_P$ and is in good agreement with $\Delta H_L^C=H_P$, suggesting that the wall forms a Neel-like structure. In contrast, $\Delta H_L^C$ is much larger than $H_P$ for the Ta and W underlayer films, indicating the presence of a Bloch-like wall. This is consistent with the relatively small $D$ found (Fig. 6(c)) for the Ta underlayer films. However, it is surprising that the W underlayer films are Bloch-like given the large $D$ estimated in this system. We consider that perhaps the transverse effective field $\Delta H_T$, which is pointing opposite to that of the other underlayer films in the underlayer thickness range in interest, may play a role in defining $\Delta H_L^C$.

**Acknowledgements**

We thank A. Thiaville, Y. Lau, C. Pai, and R. A. Buhrman for helpful discussions. This work was partly supported by the FIRST program from JSPS and the Grant-in-Aid (25706017) from MEXT.

**Author Contributions**

M.H. planned the study. J.T. and M.H. wrote the manuscript. J.S. performed film deposition and film characterization, J.T., J.S. and M.H. fabricated the devices. J.K. evaluated the current induced effective field, J.T. carried out the Kerr measurements and analyzed the data with the help of M.H., M.Y., S.M. and H.O. All authors discussed the data and commented on the manuscript.

**Competing financial interests**

The authors declare that they have no competing financial interests.

**Figure captions**

**Fig. 1. Schematic of the experimental setup and magneto-optical Kerr images illustrating current induced domain wall motion**. (a) Optical microscopy image of the wire used to study current induced domain wall motion. The Ta|Au electrodes are indicated by the yellow colored region. A pulse generator is connected to one of the Ta|Au electrodes, as schematically shown. (b) Illustration of the experimental setup. The thick black arrows indicate the magnetization of the CoFeB layer. (c,d) Typical Kerr images showing current induced domain wall motion along (c) and against (d) the electron flow for wires with different underlayers: (c) ~0.5 nm thick Ta underlayer, (d) ~3.6 nm thick TaN(Q=0.7%) underlayer. Domain walls in (c,d) are both ↓↑ walls. Between images: ~−40 V, 100 ns long pulses are applied 12 times for (c) and ~28 V, 100 ns long pulses are applied 20 times for (d).

**Fig. 2. Pulse amplitude dependence of domain wall velocity**. (a-c) Domain wall velocity as a function of pulse amplitude plotted for various underlayer thicknesses. The underlayer is (a) Hf, (b) Ta, (c) TaN(Q=0.7%) and (d) W. The direction to which the wall moves is indicated in each panel of (d): e⁻ corresponds to electron flow. Average domain wall velocity is obtained by moving a domain wall across ~30 μm long wires. This process is repeated multiple times for a given pulse amplitude: all results are shown by the symbols.

**Fig. 3. Threshold current density and propagation field required to move domain walls.** (a-h) The threshold current density needed to move a domain wall is shown by the open symbols, whereas the solid symbols represent the *maximum* current density applied to each device; beyond this current density, current induced domain nucleation occurs. The current density that flows in the CoFeB layer ($J_C^{CoFeB}$) and the underlayer ($J_C^{Under}$) are shown in the top (a-d) and bottom (e-h) panels. The current density in each layer is estimated by taking

into account the difference in the resistivity of the two layers. (i-l) Propagation field $H_P$ plotted as a function of the underlayer thickness. Solid and open symbols represent positive and negative $H_P$, respectively. The absolute value of $H_P$ is shown. The underlayer is (a,e,i) Hf, (b,f,j) Ta (c,g,k) TaN(Q=0.7%) and (d,h,l) W.

**Figure 4. Current induced effective field vs. the underlayer thickness**. Transverse (a-d) and longitudinal (e-h) components of the current induced effective field, $\Delta H_T/J^{Under}$ and $\Delta H_L/J^{Under}$, respectively, are plotted as a function of the underlayer thickness for film stacks with different underlayers: (a,e) Hf, (b,f) Ta (c,g) TaN(Q=0.7%) and (d,h) W. The effective field is normalized by the current density ($J^{Under}=1\times10^8$ A/cm$^2$) that flows into the underlayer. The solid and open symbols correspond to the effective field when the magnetization of the CoFeB layer is pointing along +Z and –Z, respectively.

**Figure 5. The offset field associated with the chiral magnetic texture**. (a,b) Domain wall velocity plotted as a function of the longitudinal field $H_L$ (directed along the current flow and the wire's long axis) for two different film stacks: (a) ~0.5 nm thick Ta underlayer and (b) ~3.6 nm thick TaN(Q=0.7%) underlayer. Blue circles and red triangles indicate the wall velocity when positive and negative voltage pulses are applied, respectively. Left (right) panel shows results for ↓↑ (↑↓) wall. Solid lines are linear fits to the data to obtain $H_L^*$. The pulse amplitude is ~±40 V for (a) and ~±28 V for (b). (c-f) The offset field $H_L^*$, i.e. the longitudinal field ($H_L$) at which the velocity becomes zero, plotted as a function of underlayer thickness. The underlayer is (c) Hf, (d) Ta, (e) TaN(Q=0.7%) and (f) W. Solid and open symbols represent ↑↓ and ↓↑ domain walls, respectively. $H_L^*$ is evaluated when the wall is driven either by positive or negative voltage pulses: here, both results are shown together.

The background color of each panel indicates the direction to which a corresponding domain wall moves; red: against the electron flow, blue: along the electron flow. Solid and dashed lines represent fitting using Eq. (1) to estimate $D$.

**Figure 6. Spin Hall angle and the interface Dzyaloshinskii-Moriya interaction strength**. (a-c) Spin Hall angle (a) and the spin diffusion length (b) plotted for different underlayer materials. The center panel shows the dependence on the $N_2$ gas concentration ($Q$, in at%) during the Ta reactive sputtering. (c) Dzyaloshinskii-Moriya exchange constant $D$ as a function of underlayer material. The center panel shows $D$ against the atomic concentration of N in TaN. (d) Threshold longitudinal effective field required to move a domain wall via spin Hall torque $\Delta H_L^C$ plotted against the propagation field $H_P$. The solid line shows $\Delta H_L^C = H_P$.

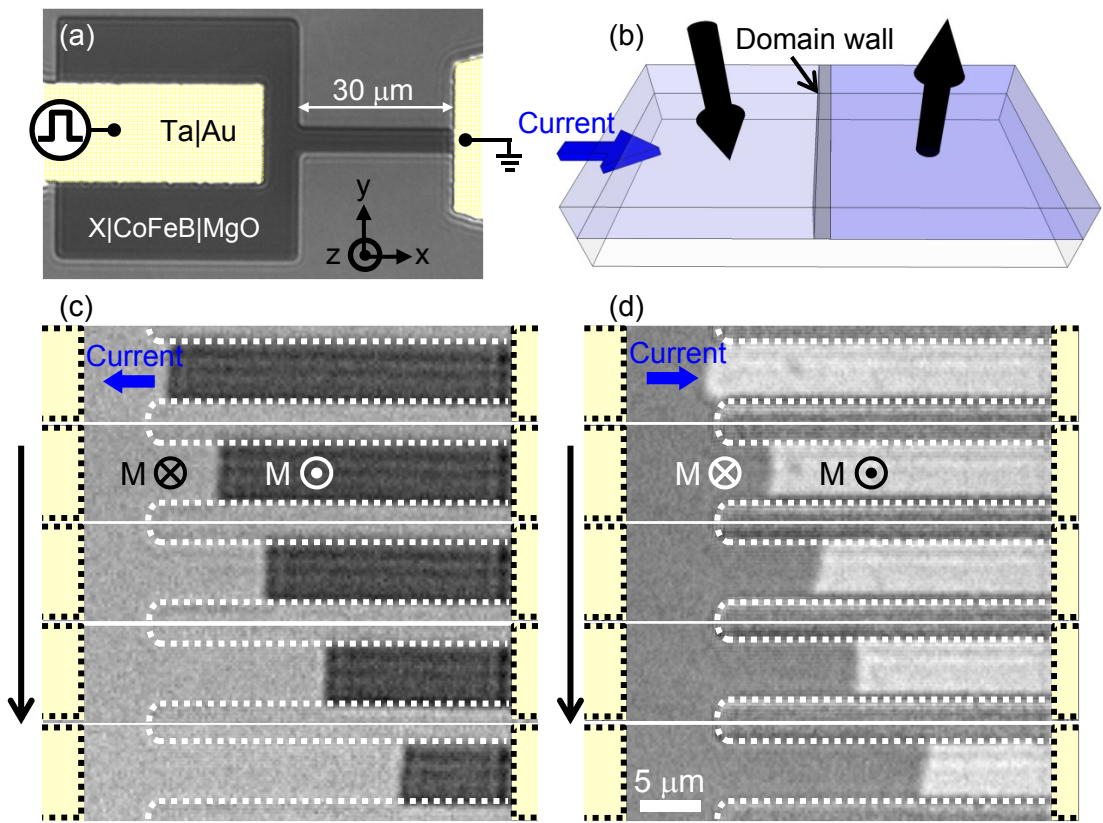

Fig. 1

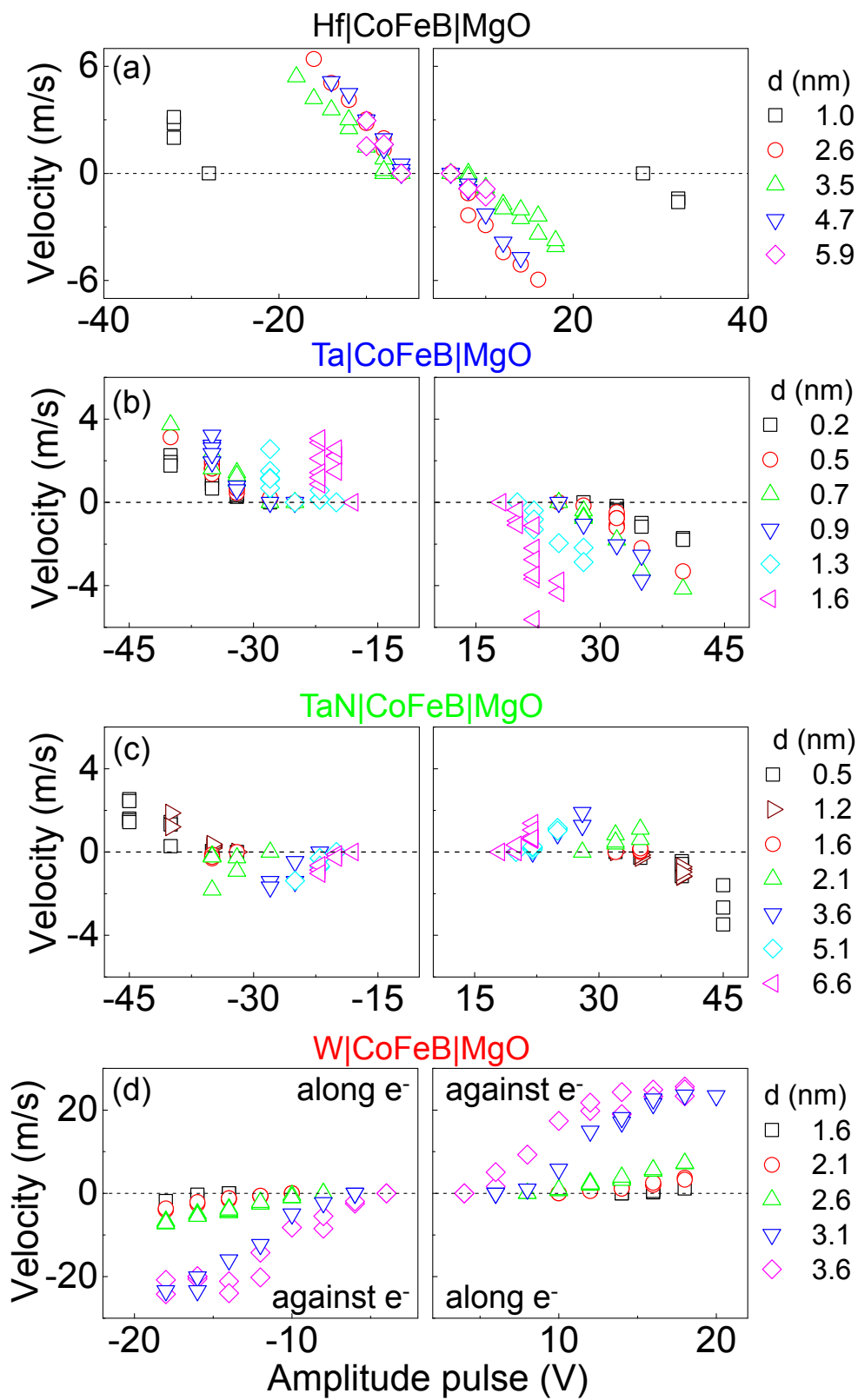

Fig. 2

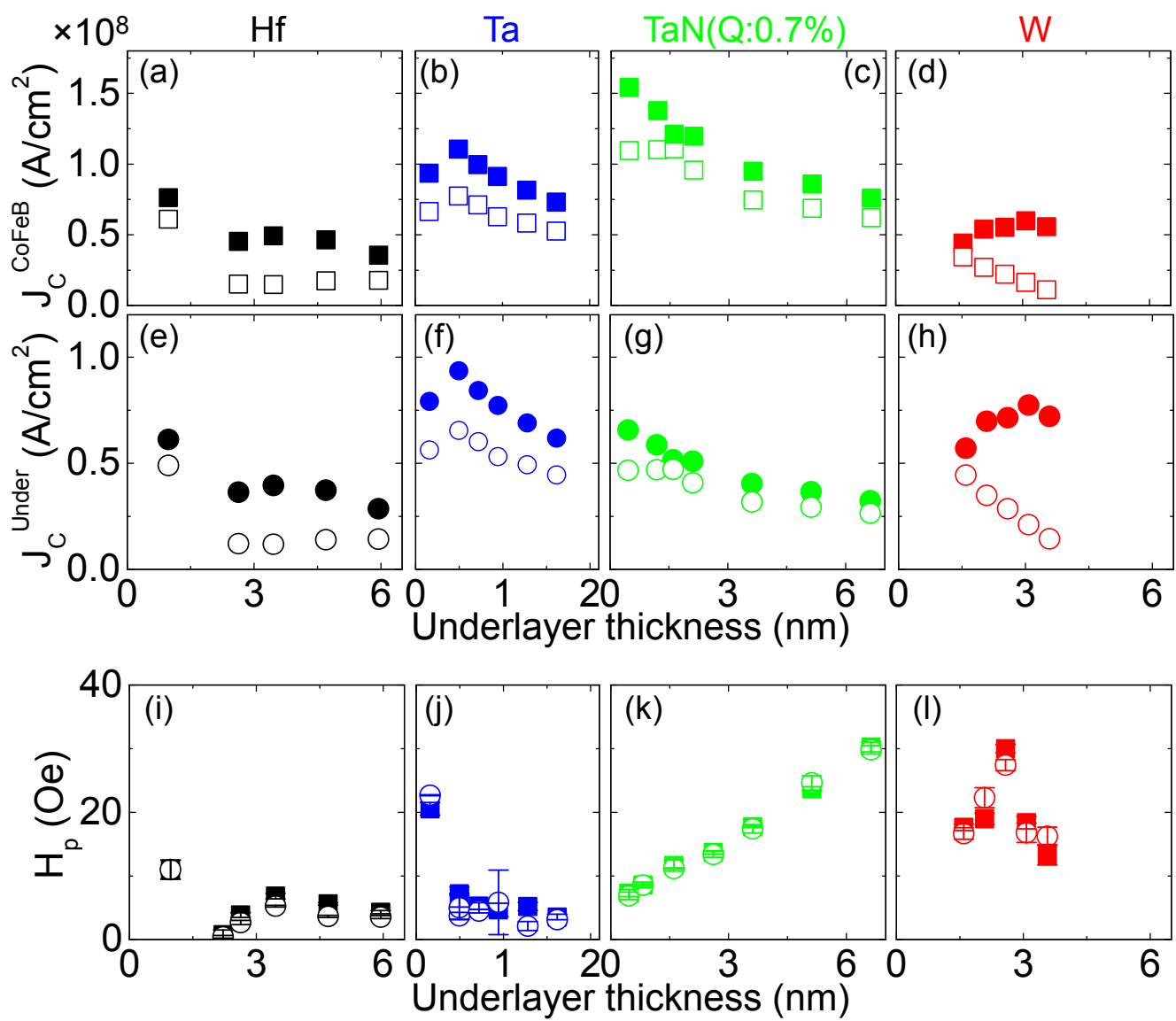

Fig. 3

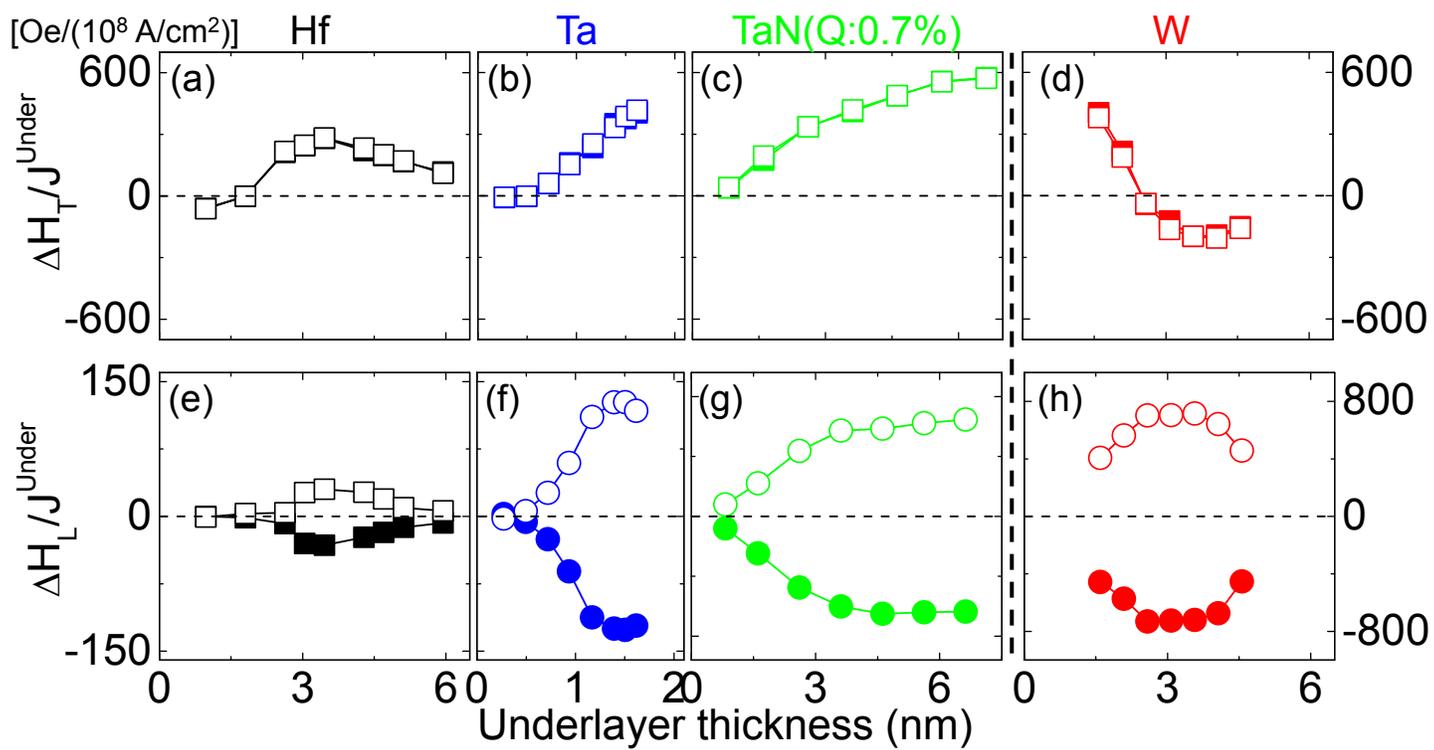

Fig. 4

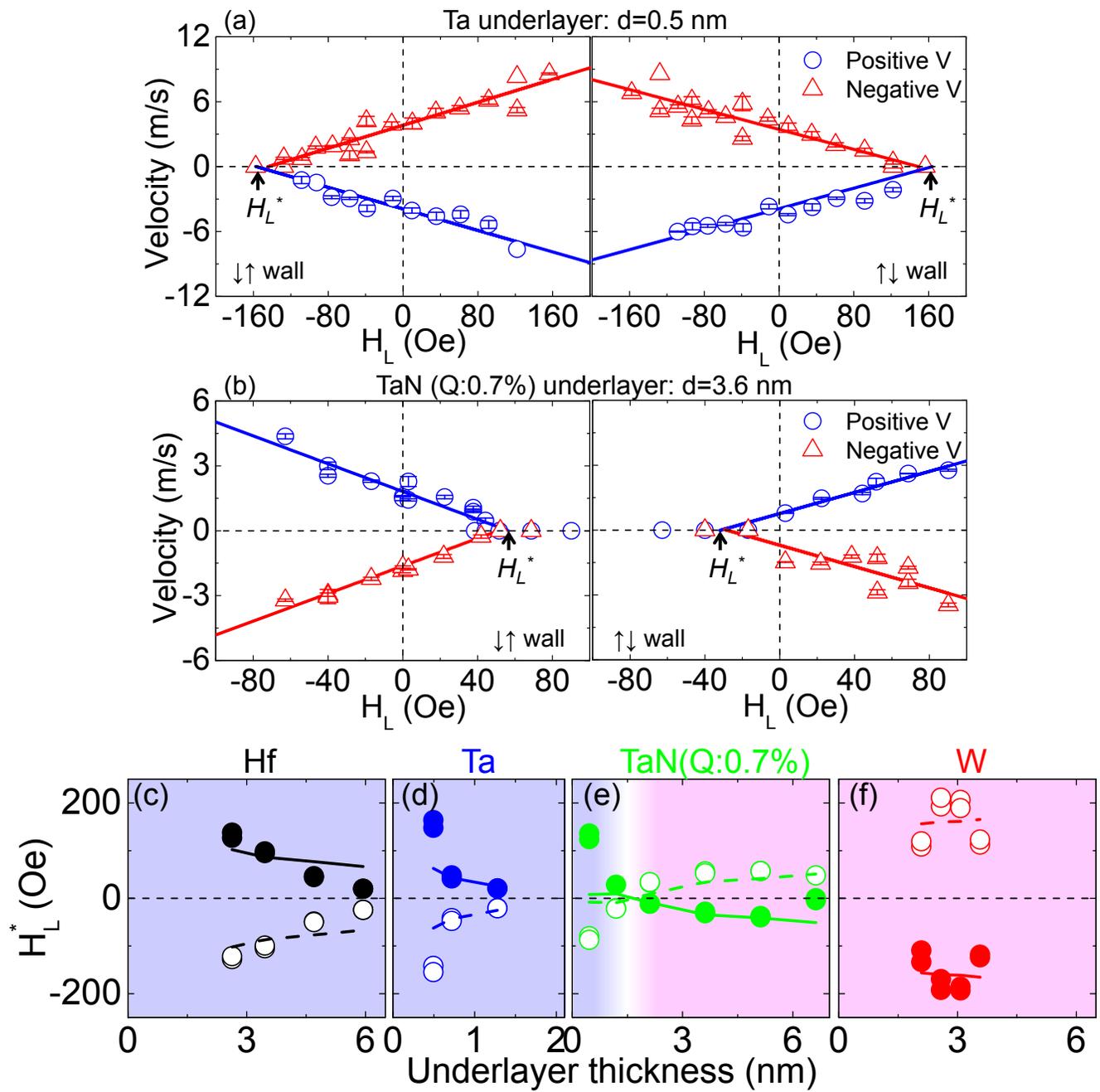

Fig. 5

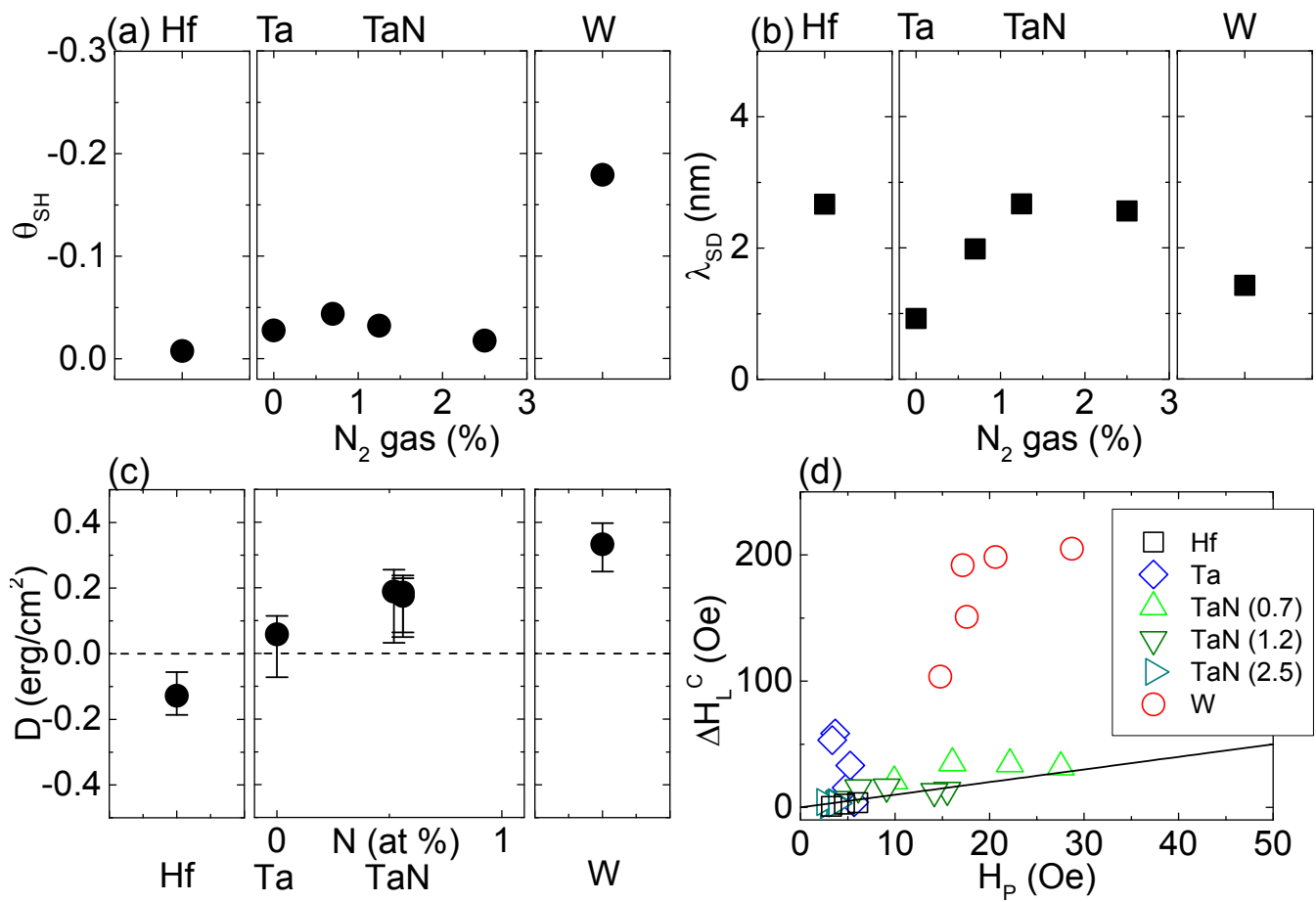

Fig. 6

**Supplementary information for:**

**Interface control of the magnetic chirality in CoFeB|MgO heterosctructures with heavy metal underlayers**

Jacob Torrejon[1], Junyeon Kim[1], Jaivardhan Sinha[1], Seiji Mitani[1] and Masamitsu Hayashi[1*]
[1]*National Institute for Materials Science, Tsukuba 305-0047, Japan*

Michihiko Yamanouchi[2,3] and Hideo Ohno[2,3,4]
[2] *Center for Spintronics Integrated Systems, Tohoku University, Sendai 980-8577, Japan*
[3]*Research Institute of Electrical Communication, Tohoku University, Sendai 980-8577, Japan*
[4]*WPI Advanced Institute for Materials Research, Tohoku University, Sendai 980-8577, Japan*

**S1. Sample preparation**

Films are deposited by magnetron sputtering (DC and RF) on Si|100 SiO$_2$ wafers. The film stack is composed of Substrate|$d$ X|1 Co$_{20}$Fe$_{60}$B$_{20}$|2 MgO|1 Ta (units in nanometer). X is Hf, Ta, TaN and W. The TaN underlayer is formed by reactively sputtering Ta in the Ar gas atmosphere mixed with a small amount of N$_2$. Ar and N$_2$ gas concentrations are controlled independently by gas mass flow meters. We define $Q$ as the atomic ratio of the N$_2$ gas over the total (Ar + N$_2$) gas, i.e. $Q \equiv \dfrac{S_{N_2}}{S_{Ar}+S_{N_2}}$, where $S_X$ denotes the mass flow (in unit of sccm) of gas X. $Q$ is varied from 0 to 2.5% here. The atomic composition of the Ta(N) films is determined by Rutherford backscattering spectroscopy (RBS): the error bar is typically ~±5%. Films are deposited using a linear shutter to vary the underlayer thickness across the wafer. The underlayer thickness $d$ is calibrated using the resistance of the patterned devices for all underlayer films. All films are post-annealed at 300 ºC for one hour in vacuum.

Magnetic properties of the films are measured using vibrating sample magnetometry (VSM). The moment per unit volume ($M/V$) and the magnetic anisotropy energy ($K_{EFF}$) are plotted in Fig. S1(a-h). Saturated moment values ($M$) are divided by the product of film area and the CoFeB thickness to obtain $M/V$. It should be noted that the nominal thickness of the CoFeB layer contains, if any, the thickness of a magnetically dead layer. Thus $M/V$ differs from the *real* saturation magnetization ($M_S$) when a dead layer is present. The magnetic anisotropy energy is estimated from the integrated difference of the out of plane and in-plane hysteresis loops. Positive $K_{EFF}$ corresponds to magnetic easy axis pointing along the film normal. For details, see Ref. [1].

Wires (for evaluating current induced domain wall motion) and Hall bars (for the analysis of current induced effective fields) are patterned by optical lithography and Ar ion etching. Subsequent lift-off process is used to make the 10 Ta|100 Au (units in nm) electrodes. Resistivity of each film is evaluated using the Hall bars. The underlayer thickness ($d$) dependence of the inverse of a normalized resistance ($1/(R_{XX} \cdot w/L)$) is plotted in Fig. S1(i-l) and Fig. S11(i-l) for all underlayer films. Average resistivity values, obtained by the linear fitting shown in Fig. S1(i-l) and Fig. S11(i-l), are tabulated in Table. 1. For W, we find a jump in the resistivity at $d \sim 5$ nm (see Fig. S1(i)), suggesting that a structural phase transition (from β-W at small $d$ to α-W at large $d$) takes place at this thickness, as reported previously[2]. Note that the magnetic anisotropy $K_{EFF}$ of W|CoFeB|MgO heterostructures (Fig. S1(h)) also drops as $d$ exceeds ~5 nm. A decrease in the resistivity is also observed for thicker Hf underlayer films, however its origin is not clear at the moment. The y-axis intercept of the linear fitting of $1/(R_{XX} \cdot w/L)$ provides the resistivity of the CoFeB layer ($\rho_{CoFeB}$). For the TaN underlayer films, which show a constant underlayer resistivity for a large $d$ range, we find $\rho_{CoFeB} \sim 160$ μΩ·cm.

Since it is difficult to estimate $\rho_{CoFeB}$ from the intercept for the other underlayer films, we assume $\rho_{CoFeB} \sim 160$ μΩ·cm throughout this paper. Separate CoFeB wedge films are made to check this assumption and we find relatively close values.

**S2. Experimental methods**

**A. Magneto-optical Kerr effect and the hysteresis loops**

Magneto-optical images of the wires are acquired using Kerr microscopy. To quantify the magnetic contrast, the region of interest (ROI, i.e. the wire) is selected in the acquired image and converted into a two dimensional arrays of integer. The average value of the Kerr intensity (i.e. the CCD signal) of the ROI, denoted as $I$ hereafter, is plotted in Fig. S2(a) as a function of the out of plane field $H_Z$. Hysteresis loops of wires with different TaN(Q=0.7%) underlayer thicknesses are shown. For the thicker underlayer films $I$ is larger when the magnetization is pointing up (large positive $H_Z$). In contrast, $I$ is larger for magnetization pointing downward for the thinner underlayer films. The difference in $I$ when the magnetization is pointing "up" and "down" is defined as $\Delta I$ and the mean value of $I$ is denoted as $I_0$. Figure S2(b) shows $\Delta I/I_0$ as a function of the underlayer thickness for the three film structures investigated here. The sign of $\Delta I/I_0$ changes at a certain thickness for each film structure.

These changes in $\Delta I/I_0$ are likely due to an optical interference effect within the sample. As the total thickness of the heterostructure is very thin, a significant amount of light passes through the heterostructure (X|CoFeB|MgO|Ta) and reaches the Si|SiO$_2$ interface (the thickness of SiO$_2$ is ~100 nm). Magneto-optical Faraday effect takes place when the light transmits through the heterostructure, whereas the Kerr effect contributes to the signal reflected at the film surface. Most of the light which have transmitted through the film reach the Si|SiO$_2$ interface and get

reflected to travel toward the heterostructure. Some fraction of the reflected light can transmit through the heterostructure (and again developing the Faraday effect) and propagate toward the CCD camera; the other fraction will get reflected at the heterostructure and again travel toward the Si|SiO$_2$ interface. This will develop an interference effect in the 100 nm thick SiO$_2$ layer and the magneto-optical signal captured with the CCD camera likely includes contribution from both the Kerr and the Faraday effects. Such multiple reflections/interference can change the size and sign of $\Delta I/I_0$. Note that we do not observe any change in the sign of $\Delta I/I_0$ when we use naturally oxidized Si substrates (with just a few nanometer thick SiO$_2$), confirming that the effect is optical (and not electronic).

All images shown in this paper are subtracted images. An image of a uniformly magnetized state with magnetization pointing along –z is captured as the reference image. This reference image is subtracted from each image.

**B. Domain wall nucleation using current pulses**

A domain wall is nucleated by applying voltage pulses to the wire. First, the CoFeB layer is uniformly magnetized by applying an out of plane field $H_z$. The field is then reduced to zero and we apply a voltage pulse (typically 100 ns of duration) to nucleate a domain wall. This process typically suffices to create one or two domain walls within the wire. In some film structures, an additional magnetic field application is required to change the domain pattern after the pulse application to form an appropriate domain structure.

**C. Propagation field of the domain walls**

The out of plane field needed to move a domain wall, i.e. the propagation field, is evaluated using Kerr microscopy images. After the domain wall nucleation process, the out of plane field $H_Z$ is ramped towards higher magnitude, either to positive or negative $H_Z$, and the magnetic state is monitored with the Kerr microscopy. Such measurement cycle is repeated in each device 10 times (5 times for positive and 5 times for negative $H_Z$). The propagation field $H_P$ is defined as the field ($H_Z$) at which the Kerr signal change exceeds 50% of the total change expected. The field ramp rate is approximately 1 Oe/sec near the propagation field.

**D. Current induced effective field measurements**

Current induced effective field is measured in the same manner as described in Ref. [3]. A Hall bar is patterned on the same wafer with the wires. To obtain the effective field, a sinusoidal constant amplitude voltage is applied to the Hall bar and the first and second harmonic Hall voltages are measured using lock-in amplifiers. The resistance does not change with the voltage within the range we apply, thus the excitation can be treated as a constant amplitude sinusoidal current. An in-plane magnetic field directed along or transverse to the current flow is applied to evaluate the longitudinal ($\Delta H_L$) and transverse ($\Delta H_T$) components of the effective field, respectively.

Contribution from the planar Hall effect[4,5] is taken into account in obtaining $\Delta H_L$ and $\Delta H_T$. The underlayer thickness dependences of the anomalous ($\Delta R_{AHE}$) and planar ($\Delta R_{PHE}$) Hall effects are shown in Fig. S3(a-d); the ratio of $\Delta R_{PHE}$ to $\Delta R_{AHE}$ is plotted in Fig. S3(e-h). The planar Hall effect is ~5% in magnitude of the anomalous Hall effect for these film structures. We use an average value of $\Delta R_{PHE}/\Delta R_{AHE}$~0.05 to calculate $\Delta H_L$ and $\Delta H_T$ for the W and Hf underlayer films.

The spin Hall angle $\theta_{SH}$ is estimated using the following the relation[6]: $\Delta H_L = \theta_{SH} \frac{\hbar}{2e} \frac{J^{Under}}{M_S t_F}$ when the underlayer thickness $d$ is much larger than its spin diffusion length $\lambda_{SD}$. Here, $\hbar$ is the Plank constant, $e$ is the electric charge, $M_S$ and $t_F$ are the saturation magnetization and the thickness of the magnetic layer, respectively, and $J^{Under}$ is the current density that flows in the underlayer. For $M_S \cdot t_F$, we take the value of $M/V$ from Fig. S1(a-d) and multiply the nominal thickness of the CoFeB layer (here it is 1 nm). The spin diffusion length $\lambda_{SD}$ is estimated by fitting an error function to the data to obtain the underlayer thickness at which $\Delta H_L$ becomes one half of its saturated value.

We next compare the spin Hall angle $\theta_{SH}$ estimated using the above relation with reports from other groups. The value ($\theta_{SH} \sim -0.18$) for W is of the same order of magnitude with that of Pai *et al.*[2] ($\theta_{SH} \sim -0.33$). However, $\theta_{SH}$ for Ta is much smaller than that from Liu *et al.*[6]: we find $\theta_{SH} \sim -0.03$ whereas Liu et al. report $\theta_{SH} \sim -0.15$. First, due to the relatively thick magnetic dead layer[1] present in our Ta underlayer films, $M/V$ is much smaller than that of bulk CoFeB. For example, $M/V$ of the CoFeB layer for $d \sim 1.5$ nm ($d$ is the Ta underlayer thickness) is $\sim 690$ emu/cm$^3$, which is $\sim 60\%$ of that reported by Liu *et al.* In addition, we find significant amount of the transverse component of the spin Hall torque (equivalent to the field-like term): $\Delta H_T$ is nearly $\sim 3$ times larger than $\Delta H_L$ when $d$ is larger than $\lambda_{SD}$, whereas Liu *et al.* find negligible sign of such component. The above relation used to estimate the spin Hall angle is only valid when the interface (here it is Ta|CoFeB) is transparent for spin transmission[7]. For transparent interface, the field-like term ($\Delta H_T$) should be zero. The large $\Delta H_T$ found in our films indicates that the interface is far from transparent and thus we need to take this into account to properly estimate the spin Hall angle. Since $\Delta H_T$ is a $\sim 4$ times smaller than $\Delta H_L$ (when $d > \lambda_{SD}$) for the W underlayer films, perhaps the estimation works better in this system.

**E. Current induced domain wall motion**

Current induced domain wall motion is studied by applying voltage pulses to the wire and the distance the domain wall traveled after the pulse application is evaluated using Kerr microscopy. Typically, 5 pulses, each 100 ns long and separated by ~10 ms, are applied to the sample and a magneto-optical image of the device is captured ~1 s later the pulse application by a CCD camera attached to the Kerr microscopy. This process is repeated multiple times to move a domain wall across the ~30 μm long wire. We fit the wall position as a function of cumulated pulse length with a linear function to obtain the wall velocity. In general, the velocity is nearly constant when the wall is moving. Once the wall gets pinned, the velocity is zero and these points are excluded from the fitting. Such processes of moving a domain wall across the wire are performed multiple times for a given pulse amplitude.

Fig. S4 shows exemplary profiles of how a domain wall propagates along the wire depending on the film structure. We find that the wall motion depends on the strength of spin Hall torque and/or spin transfer torque. If the size of spin Hall/spin transfer torque is strong enough, the wall moves along the wire relatively smoothly without pinning (Fig. S4(a)). In contrast, when the spin Hall/spin transfer torque is relatively small (for example, in TaN(Q=2.5%) underlayer films), the domain wall often gets pinned at a local pinning center (Fig. S4(b)). For such profiles, we fit the position vs. cumulated pulse length only when the wall is moving and take an average value of the slopes.

**E. Joule heating measurements**

To estimate the temperature rise due to Joule heating, real time measurements of the anomalous Hall resistance during the pulse application are performed. Given the small temperature variation of the resistivity of these films (results not shown), it is difficult to conduct a time-domain reflection/transmission measurements to infer the device temperature from the sample resistance. Here we use the anomalous Hall resistance to calibrate the device temperature[3]. Joule heating is measured in Sub|0.4 Ta|1 CoFeB|2 MgO|1 Ta since this film structure has negligible current induced effective field ($\Delta H_T$ and $\Delta H_L$ are nearly zero[3]). Presence of $\Delta H_T$ and $\Delta H_L$ will influence the estimation of Joule heating.

Figure S5(a) shows an exemplary optical microscopy image of a Hall bar device, similar in structure to what has been used for the measurements. A pulse generator and three input channels ($V_1$, $V_2$ and $V_3$) of an oscilloscope are connected to the terminals of a Hall bar. A voltage pulse is applied from the pulse generator. The hall resistance is obtained by

$$R_{XY}(t) = Z_0 \left[ \frac{V_2(t) - V_3(t)}{V_1(t) + V_2(t) + V_3(t)} \right],$$ where $Z_0$ is the characteristic impedance (50 Ω) of the system.

The anomalous hall resistance $\Delta R_{XY}$ is defined as half the difference in the Hall resistance when magnetization is directed along +Z and –Z, i.e. $\Delta R_{XY} \equiv \frac{1}{2} \left[ R_{XY} (M_z \| Z) - R_{XY} (M_z \| -Z) \right]$.

Time resolved measurement of the anomalous Hall resistance studied at low current density $J_0$ are shown in Fig. S5(b). Scope traces are averaged over ~128 times to improve the signal to noise ratio. Figure S5(c) shows DC measurement (10 μA current) of the Hall resistance $R_{XY}$ to compare with that of Fig. S5(b). $\Delta R_{XY}$ obtained by the time resolved measurements (~45 Ω) shows relatively good agreement with that of the DC measurements (note that high frequency losses are not taken into account in the real time measurements).

Differences in $\Delta R_{XY}$ when high ($J$) and low ($J_0$) current density pulses are applied to the Hall bar, i.e. $\Delta R_{XY}(J)-\Delta R_{XY}(J_0)$, are shown in Fig. S5(d) for various $J$. We find observable difference when $J$ exceeds $\sim 5\times 10^7$ A/cm$^2$. The difference $\Delta R_{XY}(J)-\Delta R_{XY}(J_0)$ continues to increase after ~100 ns: this is primarily to do with the relatively poor heat conduction of the substrate (the 100 nm thick thermal oxide (SiO$_2$) hinders the heat flow to the substrate). To estimate the device temperature, $\Delta R_{XY}(J)/\Delta R_{XY}(J_0)$ is plotted as function of $J$ in Fig. S5(e). The solid line shows fitting with a parabolic function: $\Delta R_{XY}(J)/\Delta R_{XY}(J_0)=aJ^2+1$ where $a$ is a fitting parameter. Previously, we have measured the change in the DC anomalous Hall resistance with temperature in the corresponding device[3]. Using this relation and the results shown in Fig. S5(e), we estimate the device temperature as function of $J$ when a 100 ns long pulse is applied. The estimates are shown in Fig. S5(f): the solid line indicates a parabolic fitting. We find that the temperature rise is ~200 K above room temperature for the maximum current density applied here.

## S3. One dimensional model of a domain wall

### A. Modified Landau Lifshitz Gilbert equation

The one dimensional (1D) model[8] describing domain wall dynamics is used to understand the effect of the spin Hall effect, the DMI and the spin transfer torque. The dynamics of a domain wall is described by two parameters, its position $q$ and magnetization angle $\Psi$. For out of plane magnetized samples, the domain wall magnetization points along a direction within the film plane: we define $\Psi=0$ and $\pi$ corresponding to the Bloch wall and $\Psi=\pi/2$ and $3\pi/2$ as the Neel wall. The two coupled equations that describe the dynamics of $(q, \Psi)$ read:

$$(1+\alpha^2)\frac{\partial q}{\partial t} = \left[\frac{1}{2}\gamma\Delta H_K \sin 2\psi + \frac{\pi}{2}\gamma\Delta H_T \sin\psi - \frac{\pi}{2}\gamma\Delta\left(H_L + \Gamma H_{DM}\right)\cos\psi + u\right]$$
$$+ \alpha\left[-\frac{\gamma\Delta}{2M_S}\left(\frac{\partial \sigma_{PIN}}{\partial q}\right) + \gamma\Delta\Gamma H_Z + \gamma\Delta\Gamma H_{SH}\sin\psi + \beta u\right]$$
(S1a)

$$(1+\alpha^2)\frac{\partial \psi}{\partial t} = -\alpha\left[\frac{1}{2}\gamma H_K \sin 2\psi + \frac{\pi}{2}\gamma H_T \sin\psi - \frac{\pi}{2}\gamma(H_L + \Gamma H_{DM})\cos\psi + \frac{u}{\Delta}\right]$$
$$+\left[-\frac{\gamma\Delta}{2M_S}\left(\frac{\partial \sigma_{PIN}}{\partial q}\right) + \gamma\Gamma H_Z + \gamma\Gamma H_{SH}\sin\psi + \beta\frac{u}{\Delta}\right] \tag{S1b}$$

Here, $\gamma$ is the gyromagnetic ratio, $H_K$ is the magnetic anisotropy associated with the domain wall magnetization, $\Delta$ is the domain wall width parameter and $\alpha$ is the Gilbert damping constant. The effect of a pinning potential is described by the term with $\partial\sigma_{PIN}/\partial q$: $\sigma_{PIN}$ denotes the wall pinning potential energy density. Spin transfer torque is represented by $u = -\frac{\mu_B P}{eM_S}J$, where $\mu_B$ and $e$ are the Bohr magnetron and the electron charge (we define $e>0$ for convenience), $P$ and $M_S$ are the current spin polarization and saturation magnetization of the ferromagnetic material and $J$ is the current density that flows through the magnetic (CoFeB) layer. $\beta$ is the non-adiabatic spin torque term[9]. $H_Z$, $H_T$ and $H_L$ correspond to the out of plane, in-plane transverse (transverse to the wire's long axis) and in-plane longitudinal (along the wire's long axis) fields, respectively. $\Gamma$ represents the domain pattern; $\Gamma=+1$ for ↑↓ wall and $\Gamma=-1$ for ↓↑ wall. The spin Hall torque is modeled[10] by an effective out of plane magnetic field $H_{SH}\sin\psi$. $H_{SH}$ is equivalent to $\Delta H_L$ in the main text. The DMI is included as an effective offset in-plane field $\Gamma H_{DM}$ directed along the wire's long axis[10-12]. For a domain wall spiral, the offset field $H_{DM}$ changes its sign depending on the domain pattern $\Gamma$ (that is, ↑↓ or ↓↑ walls).

To describe experimental results using Eq. (S1a,b), we introduce the following parameter. Since $u$ and the spin Hall effective field scales with the current density, we define $u = -\tilde{P}j$ and $H_{SH} = -\tilde{\theta}_{SH}j$. Here $\tilde{P} \equiv \frac{\mu_B P|J|}{eM_S}$, $\tilde{\theta}_{SH} \propto \theta_{SH}\frac{\pi}{2}\frac{\hbar|J|}{2eM_S t_F}$, $\theta_{SH}$ is the spin Hall angle, $t_F$ is the

thickness of the magnetic layer and $j$ represent the direction of current, i.e. $j = \frac{J}{|J|}$. For simplicity, we use the same notation $J$ for current that flows in the magnetic layer ($J^{CoFeB}$) and the underlayer ($J^{Under}$), if not explicitly stated otherwise. The sign of the spin Hall angle is set as the following: $\theta_{SH} > 0$ for Pt and $\theta_{SH} < 0$ for Ta. The chirality of the domain wall spiral is denoted by $\chi$: $\chi=1$ for right handed and $\chi=-1$ for left handed domain walls. The Dzyaloshinskii-Moriya offset field $H_{DM}$ ($=\frac{D}{M_S \Delta}$, $D$ is the Dzyaloshinskii-Moriya exchange constant) can then be expressed using $\chi$ as: $H_{DM} = \chi |H_{DM}|$. Substituting these parameters into Eq. (S1a) and (S1b) gives:

$$(1+\alpha^2)\frac{\partial q}{\partial t} = \left[ \frac{1}{2}\gamma \Delta H_K \sin 2\psi + \frac{\pi}{2}\gamma \Delta H_T \sin\psi - \frac{\pi}{2}\gamma \Delta \left( H_L + \Gamma \chi |H_{DM}| \right) \cos\psi - \tilde{P}j \right]$$
$$+ \alpha \left[ -\frac{\gamma \Delta}{2M_S}\left( \frac{\partial \sigma_{PIN}}{\partial q} \right) + \gamma \Delta \Gamma H_z - \gamma \Delta \Gamma \tilde{\theta}_{SH} j \sin\psi - \beta \tilde{P}j \right] \quad \text{(S2a)}$$

$$(1+\alpha^2)\frac{\partial \psi}{\partial t} = -\alpha \left[ \frac{1}{2}\gamma H_K \sin 2\psi + \frac{\pi}{2}\gamma H_T \sin\psi - \frac{\pi}{2}\gamma \left( H_L + \Gamma \chi |H_{DM}| \right) \cos\psi - \frac{\tilde{P}j}{\Delta} \right]$$
$$+ \left[ -\frac{\gamma \Delta}{2M_S}\left( \frac{\partial \sigma_{PIN}}{\partial q} \right) + \gamma \Gamma H_z - \gamma \Gamma \tilde{\theta}_{SH} j \sin\psi - \beta \frac{\tilde{P}j}{\Delta} \right] \quad \text{(S2b)}$$

Left handed walls: ↑←↓ wall: $\chi=-1$, $\Gamma=1$, ↓→↑ wall: $\chi=-1$, $\Gamma=-1$

Right handed walls: ↑→↓ wall: $\chi=1$, $\Gamma=1$, ↓←↑ wall: $\chi=1$, $\Gamma=-1$

$\tilde{P} > 0$ for positively spin polarized materials (e.g. Py, Co, CoFeB)

$\tilde{\theta}_{SH} > 0$ for Pt, Pd, etc., $\tilde{\theta}_{SH} < 0$ for Hf, Ta, W, etc.

$j=+1$ for current flowing along +x, $j=-1$ for current flowing along –x.

## B. Domain wall velocity under spin Hall and spin transfer torques

When $\Psi$ is small and $\sigma_{PIN}=0$ (no pinning), Eqs. (S2a) and (S2b) can be linearized to calculate the domain wall velocity (below the Walker breakdown limit). The solution is given as:

$$v_{DW} \equiv \frac{\partial q}{\partial t} = \frac{\pm\gamma\Delta\frac{\pi}{2}\Gamma\tilde{\theta}_{SH}j}{-\Gamma\tilde{\theta}_{SH}j+\alpha\left(\mp H_K-\frac{\pi}{2}H_T\right)}\left(H_L+\Gamma\chi|H_{DM}|\right)$$
$$-\frac{-\Gamma\tilde{\theta}_{SH}j+\beta\left(\mp H_K-\frac{\pi}{2}H_T\right)}{-\Gamma\tilde{\theta}_{SH}j+\alpha\left(\mp H_K-\frac{\pi}{2}H_T\right)}\tilde{P}j+\gamma\Delta\frac{\left(\mp H_K-\frac{\pi}{2}H_T\right)}{-\Gamma\tilde{\theta}_{SH}j+\alpha\left(\mp H_K-\frac{\pi}{2}H_T\right)}\Gamma H_Z$$

(S3)

where the upper and lower (plus/minus) signs indicate cases for $\Psi$ close to 0 and $\pi$, respectively.

The domain wall velocity is calculated numerically and plotted in the lower and upper panels of Fig. S6(a-d) for cases with and without contribution from the spin transfer torque (STT), respectively. Here we assume $\tilde{\theta}_{SH}<0$, as in all of the underlayers studied here. The blue fat arrow in Fig. S6(a-d) represents the current flow. The effective out of plane field due to spin Hall torque is illustrated by the red fat arrow when positive current (+I) is applied. The direction to which a domain wall moves (for +I) with spin Hall torque and STT are shown by the red and orange thin arrows, respectively. The velocity is calculated for each domain pattern ($\Gamma$: ↑↓ and ↓↑ walls) and each chirality ($\chi$: left or right handed).

Without STT, the field at which the velocity becomes zero, defined as $H_L^*$ similar to the experiments, coincides with the offset field $H_{DM}$. However, as STT is turned on, $H_L^*$ deviates from $H_{DM}$. For left handed walls, STT increases the magnitude of the wall velocity at zero field and the magnitude of $H_L^*$. In contrast, for right handed walls, the zero field velocity and $H_L^*$ both decreases in magnitude when STT is added. $H_L^*$ can even change its sign, as shown in Fig. S6(c,d), when the STT contribution becomes larger than that of the spin Hall torque. $H_L^*$ varies with STT since STT changes the velocity at zero $H_L$ and consequently shifts the $v_{DW}$ vs. $H_L$ vertically (see Fig. S6). These results are consistent with Eq. (S3), from which we find (assuming $H_Z=H_T=\beta=0$):

$$H_L^* = -H_{DM} \mp \frac{2}{\pi}\frac{u}{\gamma\Delta} = -\Gamma\left[D + \text{sgn}(\theta_{SH})\frac{2}{\pi}\frac{\mu_B P}{\gamma e}|J^{CoFeB}|\right]\frac{1}{M_S\Delta} \tag{S4}$$

The minus/plus sign in the expression after the first equality sign, corresponding to $\Psi$ close to $\pi$ or 0, respectively, is determined by the sign of the spin Hall angle, the direction of current and the domain pattern ($\Gamma$). Equation (S4) reduces to Eq. (1) in the main text if all constants are substituted.

The experimentally obtained underlayer thickness ($d$) dependence of the offset field $H_L^*$ is fitted using Eq. (S4). Results are shown by the solid and dashed lines in Fig. 5(c-f). We vary $P$ from 0 to 1 to study the contribution from spin transfer torque ($P=0$ corresponds to the case when spin transfer torque is absent). The $d$ dependence of $K_{EFF}$, shown in Fig. S1(e-h), is used to estimate the domain wall width $\Delta = \sqrt{A/K_{EFF}}$. We use $A$=3.1 erg/cm$^2$ obtained from a separate study in a similar film structure[13]. The saturation magnetization ($M_S$) is assumed to be ~1600 emu/cm$^3$, which is larger than that of bulk $Co_{20}Fe_{60}B_{20}$ but smaller than that of bulk $Co_{25}Fe_{75}$. Our previous study on $M_S$ in CoFeB|MgO heterostructures with Ta and TaN underlayers indicate that the $M_S$ lies between these values[1]. $J^{CoFeB}$ is obtained by considering the resistivity and thickness differences between the CoFeB and the underlayer.

**C. Thermally activated domain wall motion**

The effect of pinning and thermally activated motion on $H_L^*$ is studied. STT is turned off and we study whether $H_L^*$ evaluated using the methods described in the experiments deviates from $H_{DM}$ due to the presence of pinning (in the absence of STT, $H_L^*$ should coincide with $H_{DM}$, as described above). A sinusoidal pinning potential $\sigma_{PIN} = \frac{Vq_0}{2\pi}\left[1-\cos\left(\frac{2\pi}{q_0}q\right)\right]$ (Fig. S7(a)) is introduced in the model to mimic the pinning profile along the wire. The effect of thermal activation is modeled by adding Langevin random field and allowing the initial position ($q$) and magnetization angle ($\Psi$) to vary according to normal (Gaussian) distribution[14]. The coupled equations (S2a) and (S2b) are solved ten to fifty times with different initial conditions for each parameter. The symbols and the error bars represent the mean and standard deviation of the velocity obtained from the calculations. Figure S7(b) shows the depinning probability as a function of magnetic field applied along the film normal. The device temperature is varied to show the thermally activated depinning process. The depinning field (or the propagation field) drops as the temperature is raised.

The effect of pinning on the wall velocity is shown in Fig. S7(d). Figures S7(d) and (c), open symbols, show the domain wall velocity vs. the longitudinal in-plane field ($H_L$) for wires with and without the pinning, respectively. When pinning is introduced, there is a range of $H_L$ in which the wall does not move or moves very slowly (defined as the "pinning regime" hereafter). This pinning regime is centered at $H_L \sim H_{DM}$. The size of this pinning regime expands when the strength of pinning is increased (results not shown).

The effect of thermal activation is shown by the solid symbols and the error bars in Fig. S7(d). The device temperature is set to 500 K. Near the pinning regime, the velocity distribution becomes large, as shown by the large error bars. In all cases, however, when $H_L^*$ is estimated by fitting the wall velocity outside the pinning regime using a linear function, the intersection of the linear fitting and the x-axis, i.e. $H_L^*$, coincides with $H_{DM}$ (here it is −50 Oe). Thus in the 1D model, the impact of pinning and thermally activated motion on evaluating $H_{DM}$ can be minimized if one choose the fitting range of velocity vs. $H_L$ appropriately.

However, we find that the 1D model does not exactly describe the experimental results. In Fig. S8, we show exemplary plot of the velocity vs. $H_L$, in which we see the domain wall moving in the opposite direction at large $H_L$ ($|H_L|$>60 Oe) compared to its zero field motion. In contrast to what the 1D model predicts, the velocity across the pinning regime cannot be described by a single linear line, but rather the two lines fitting the opposite velocities seem to be shifted in $H_L$ (here by ~±30 Oe). Further investigation is required to describe this effect. From the data and the 1D model, however, one can typically interpret $H_L^*$ obtained by fitting the velocity outside the pinning regime with a linear function as the *lower limit*, in magnitude, of the offset field $H_{DM}$ (see Fig. S8 for the difference of $H_L^*$ and $H_{DM}$ when pinning is present).

### D. Threshold current

The threshold current required to depin and move a domain wall with spin Hall/spin transfer torque is numerically calculated and compared with analytical solutions. Sinusoidal pinning and thermally activated motion (device temperature is 500 K) are introduced to mimic experimental conditions.

(a) Spin Hall torque

Figure S9 (a) shows the threshold spin Hall (out of plane) effective field $H_{SH}$, termed as $\Delta H_L^C$ to employ the same notation with Fig. 6(d), as a function of the offset field $H_{DM}$. The wall is a ↑↓wall ($\Gamma$=1), thus positive and negative $H_{DM}$ correspond to right ($\chi$=1) and left ($\chi$=−1) handed walls, respectively. The wall configuration (Bloch or Nell) is defined by the relative magnitude of the offset field $H_{DM}$ and the domain wall anisotropy field $H_K$. If $|H_{DM}|$ is larger than $\sim 2H_K/\pi$, the domain wall is a Neel type ($\Psi = \pi/2$ or $3\pi/2$); otherwise, it forms an intermediate state between a Neel and a Bloch type wall. $H_{DM}$ =0 corresponds to a Bloch wall ($\Psi$=0 or $\pi$). The vertical dotted lines in Fig. S9(a) illustrate the transition between Bloch-like and Neel walls.

The numerical calculations (symbols) show that $\Delta H_L^C$ depends on $H_{DM}$ when the wall is not a Neel wall: $\Delta H_L^C$ increases as the wall becomes Bloch-like. The blue solid line displays an analytical expression of $\Delta H_L^C$ given by the 1D model (the solution is for 0 K) and describes well the numerical results. In contrast, when the domain wall becomes a Neel wall as $H_{DM}$ exceeds $2H_K/\pi$, $\Delta H_L^C$ is independent of $H_{DM}$. Analytical solution of the 1D model in this regime, shown by the red dashed lines in Fig. S9(a), dictates that $\Delta H_L^C$ is equal to the propagation field ($H_P$). Thus when $\Delta H_L^C$ is larger than the propagation field $H_P$, the model predicts that the wall is not a Neel wall. The difference between $\Delta H_L^C$ and $H_P$ provides information of the wall type.

(b) Spin transfer torque

Similarly, the threshold current for moving a domain wall with spin transfer torque can be calculated as a function of $H_{DM}$. The threshold current density is represented by the minimum $u$ needed to trigger wall motion ($u_C = -\dfrac{\mu_B P}{eM_S} J_C^{CoFeB}$). In Ta|CoFeB|MgO heterostructures with thicker CoFeB layer compared to what we use here, it has been reported that the domain wall is

driven predominantly by the adiabatic spin transfer torque[15] and that contribution from the non-adiabatic torque is small. Thus $\beta=0$ is assumed in the calculations.

The symbols in Fig. S9(b) show the numerical results; the blue solid lines indicate the analytical solution. In contrast to the spin Hall torque driven wall motion, $|u_C|$ increases with $|H_{DM}|$. Note that for adiabatic STT driven domain wall motion[16], $u_C$ is determined by $H_K$ and $\Delta$ and does not depend on the pinning strength $H_P$. The effect of $H_{DM}$ can be regarded as an in-plane magnetic field that increases the Walker breakdown instability limit, which determines $u_C$.

These results show that even when spin Hall torque and STT both work to move domain walls along the electron flow, as in the Hf underlayer films, a competition exists between the size of the favorable DMI offset field ($H_{DM}$). Spin Hall torque requires large $H_{DM}$ to force the walls to be a Neel wall, whereas STT prefers smaller $|H_{DM}|$ to avoid increasing the threshold current ($u_C$).

## S4. Supplementary experimental data

### A. The slope of DW velocity vs longitudinal field

The domain wall velocity ($v_{DW}$) linearly scales with the in-plane longitudinal field $H_L$ (see Fig. 5(a,b)). The slope ($v_{DW}/H_L$) is plotted in Fig. S10 and Fig. S15 as a function of the underlayer thickness $d$. The slope $v_{DW}/H_L$ depends on the magnetic anisotropy ($K_{EFF}$) via changes in the domain wall width, which scales with $1/K_{EFF}^{0.5}$ (see Eq. (S3)). This trend is evident in the $d$ dependence of $v_{DW}/H_L$, as shown Fig. S10 and Fig. S15; see Fig. S1 for the variation of $K_{EFF}$ with $d$.

### B. Domain wall characteristics vs. N concentration in the TaN underlayer films

The underlayer thickness dependence of the magnetic moments ($M/V$) and magnetic anisotropy ($K_{EFF}$) are shown in Fig. S11 for the Ta and TaN underlayer films. Domain wall velocity vs. the pulse amplitude for TaN(Q=1.2%) and TaN(Q=2.5%) underlayer films are shown in Fig. S12. The underlayer thickness dependence of $J_C^{CoFeB}$, $J_C^{Under}$ and $H_P$ are displayed in Fig. S13, $\Delta H_T/J^{Under}$ and $\Delta H_L/J^{Under}$ are shown in Fig. S14 and $H_L^*$ and $v_{DW}/H_L$ are presented in Fig. S15.

**Figure captions**

**Figure S1. Magnetic and transport properties of X|CoFeB|MgO heterostructures.** The saturated magnetic moment per unit volume $M/V$ (a-d), the magnetic anisotropy energy $K_{EFF}$ (e-h) and the inverse of resistance ($R_{XX}$) times wire width ($w$) divided by wire length ($L$) (i-l) are plotted as a function of the underlayer thickness for different underlayers (X) noted at the top of each panel. Inset of (l) shows $1/(R_{XX} \cdot w/L)$ for the W underlayer films for a larger $d$ range in which the structural phase transition at $d \sim 5$ nm is observed. Inset of (l) shows a wider range of $d$ and $1/(R_{XX} \cdot w/L)$.

**Figure S2: Magneto-optical properties of the heterostrucutres.** (a) Out of plane hysteresis loops measured using the Kerr microscopy for wires with TaN(Q=0.7%) underlayer; loops with various underlayer thicknesses are shown. The y-axis indicates the average CCD intensity $I$ of the wire. Each plot is shifted vertically for clarity. (b) Change in the CCD intensity with the out of plane field $\Delta I/I_0$ is plotted as a function of the underlayer thickness for films with different underlayers. Positive (negative) $\Delta I/I_0$ represents bright (dark) contrast for magnetization pointing "up" (along +z).

**Figure S3. Anomalous and planar Hall effects.** The underlayer thickness dependences of the anomalous (solid symbols) and planar (open symbols) Hall effects are plotted in (a-d). The anomalous ($\Delta R_{AHE}$) and planar ($\Delta R_{PHE}$) Hall resistances are obtained by measuring the change in the Hall resistance when the field is swept along the film normal or rotated within the film plane, respectively. (e-h) Ratio of the planar Hall to anomalous Hall resistances $\Delta R_{PHE}/\Delta R_{AHE}$ is plotted as a function of the underlayer thicknesses for films with different underlayers.

**Figure S4. Current induced domain wall motion along the wire: Position vs. cumulative pulse length.** Evlolution of the position of a domain wall when voltage pulses are applied to the wire. (a) Sub|0.5 Ta|1 CoFeB|2 MgO|1 Ta, pulse amplitude is −40 V. (b) Sub|5.3 TaN(Q=2.5%)|1 CoFeB|2 MgO|1 Ta, pulse amplitude is +20 V. Pulse length is fixed to 100 ns. The red solid lines are guide to the eye.

**Figure S5. Estimation of Joule heating.** (a) Optical microscopy image of a Hall bar similar to what have been used for measurements. Schematics of the measurement configuration are shown. (b) Time resolved measurement of the anomalous Hall resistance measured at low current density $J_0$. (c) DC measurement (10 µA current) of the Hall resistance $R_{XY}$. (d) Difference in $\Delta R_{XY}$ when high ($J$) and low ($J_0$) current densities are applied to the Hall bar for various $J$. The dashed line is a guide to the eye. (e) $\Delta R_{XY}(J)/\Delta R_{XY}(J_0)$ plotted as function of $J$. The solid line shows fitting with a parabolic function. (f) Estimated device temperature as function of $J$. The solid line indicates a parabolic fitting.

**Figure S6. Numerical calculations using the 1D model: STT vs. spin Hall torque.** (a-d) Numerically calculated domain wall velocity plotted as a function of an in-plane longitudinal field $H_L$ (along the current flow) for left handed ($\chi=-1$) ↓↑ (a), ↑↓ (b) walls and right handed ($\chi=+1$) ↓↑ (c), ↑↓ (d) walls. Calculations are shown for cases with (bottom panel) and without (top panel) the spin transfer torque (STT). Squares and circles show numerical calculations for +I ($j=+1$) and –I ($j=-1$), respectively. The solid lines indicate the analytical solution (Eq. (S3)).

The black thick arrows in each cartoon show the magnetization direction including that of a domain wall. The red fat arrow indicates the spin Hall effective field $\Delta H_L$ when current (+I) is applied. The direction to which a domain wall moves with spin Hall torque and spin transfer torque are indicated by the red and orange thin arrows, respectively.

Values used in the calculations are: $\sigma_{PIN}=0$, $H_K=200$ Oe, $\Delta=10$ nm, $\alpha=0.05$, $\beta=0$, $|H_{DM}|=50$ Oe, $\tilde{\theta}_{SH}=-10$ Oe, $\tilde{P}=0$ m/s (top panels) and 20 m/s (bottom panels). (a) $\Gamma=-1$, $\chi=-1$, (b) $\Gamma=+1$, $\chi=-1$, (c) $\Gamma=-1$, $\chi=+1$, (d) $\Gamma=+1$, $\chi=+1$.

**Figure S7. Numerical calculations using the 1D model: thermally activated propagation.** (a) Profile of the sinusoidal pinning. (b) Probability of depinning a domain wall. The wall is defined as "depinned" when it moved from its initial position by more than the width of the pinning potential ($q_0$) after the field is turned off. Calculations with different initial conditions and random field that scales with $\sqrt{T}$ ($T$: temperature) are carried out fifty times. (c,d) Domain wall velocity as a function of an in-plane longitudinal field $H_L$ for right handed ↑↓ wall: (c) no pinning and (d) sinusoidal pinning. Squares and circles show numerical calculations for +I ($j=+1$) and –I ($j=-1$), respectively. The solid lines in (c) indicate linear fitting to the numerical calculations and the same lines are replotted in (d). Device temperature is varied in (d): open symbols are for 0 K and solid symbols are for 500 K. Values used in the calculations are: $H_K=200$ Oe, $\Delta=10$ nm, $\alpha=0.05$, $\beta=0$, $|H_{DM}|=50$ Oe, $\tilde{\theta}_{SH}=-10$ Oe, $\tilde{P}=0$ m/s, $V=8000$ erg/cm$^3$, $q_0=20$ nm, $\Gamma=+1$, $\chi=+1$.

**Figure S8. The offset field associated with the chiral magnetic texture**. (a,b) Domain wall velocity plotted as a function of the longitudinal field $H_L$ for 3.5 TaN(Q=1.2%)|1 CoFeB|2 MgO|1 Ta (nm). Blue circles and red triangles indicate the wall velocity when positive and negative voltage pulses are applied, respectively. Left (right) panel shows results for ↓↑ (↑↓) wall. The pulse amplitude is ~±28 V. Lines are guide to the eye.

**Figure S9. Numerical calculations using the 1D model: threshold current.** (a,b) Threshold spin Hall (out of plane) effective field $\Delta H_L^C$ (a) and $u_C$ (b) as a function of the DMI offset field $H_{DM}$. Open and solid symbols represent positive and negative $\Delta H_L^C$ or $u_C$. Blue solid and red dashed lines indicate analytical solutions, as denoted by the colored texts. The vertical dotted lines represent the boundary between a Neel wall and an intermediate state between Neel and Bloch walls. When $H_{DM} = 2H_K/\pi$, such transition takes place. The threshold current is defined as the $\Delta H_L$ or $u$ when the terminal wall velocity exceeds 1 m/s. A sinusoidal pinning is introduced and the device temperature is set to 500 K to mimic experimental conditions. Values used in the calculations are: $H_K$=200 Oe, $\Delta$=10 nm, $\alpha$=0.05, $\beta$=0, $|H_{DM}|$=50 Oe, $V$=8000 erg/cm$^3$, $q_0$=50 nm, $\Gamma$=+1.

**Figure S10. The slope of velocity vs. longitudinal field**. (a-d) Slope of domain wall velocity ($v_{DW}$) versus $H_L$ ($v_{DW}/H_L$) is plotted as a function of underlayer thickness for film stacks with different underlayers: (a) Hf, (b) Ta (c) TaN(Q=0.7%) and (d) W. Solid and open symbols represent ↑↓ and ↓↑ domain walls, respectively. $v_{DW}/H_L$ is evaluated when the wall is driven either by positive or negative voltage pulses: here, both results are shown together. The background color of each panel indicates the direction to which a corresponding domain wall moves; red: against the electron flow, blue: along the electron flow.

**Figure S11. Magnetic and transport properties of TaN|CoFeB|MgO heterostructures.** The saturated magnetic moment per unit volume $M/V$ (a-d), the magnetic anisotropy energy $K_{EFF}$ (e-h) and the inverse of resistance ($R_{XX}$) times wire width ($w$) divided by wire length ($L$) (i-l) are plotted as a function of the underlayer thickness for TaN underlayers with different $N_2$ gas concentration ($Q$). $Q=0$ corresponds to Ta.

**Fig. S12. Pulse amplitude dependence of domain wall velocity of TaN|CoFeB|MgO heterostructures**. Domain wall velocity as a function of pulse amplitude plotted for TaN underlayer thicknesses. The underlayer is (a) TaN($Q=1.2\%$) and (b) TaN($Q=2.5\%$). The direction to which the wall moves is indicated in each panel of (b): $e^-$ corresponds to electron flow. Average domain wall velocity is obtained by moving a domain wall across ~20-30 μm long wires. This process is repeated multiple times for a given pulse amplitude: all results are shown by the symbols.

**Fig. S13. Threshold current density and propagation field required to move domain walls for TaN|CoFeB|MgO heterostructures.** (a-h) The threshold current density needed to move a domain wall is shown by the open symbols, whereas the solid symbols represent the *maximum* current density applied to each device; beyond this current density, current induced domain nucleation occurs. The current density that flows in the CoFeB layer ($J_C^{CoFeB}$) and the underlayer ($J_C^{Ta(N)}$) are shown in the top (a-d) and bottom (e-h) panels. The current density in each layer is

estimated by taking into account the difference in the resistivity of the two layers. (i-l) Propagation field $H_P$ plotted as a function of the underlayer thickness. Solid and open symbols represent positive and negative $H_P$, respectively. (The absolute value of $H_P$ is shown.) The underlayer is (a,e,i) Ta, (b,f,j) TaN(Q=0.7%) (c,g,k) TaN(Q=1.2%) and (d,h,l) TaN(Q=2.5%).

**Figure S14. Current induced effective field vs. the underlayer thickness for TaN|CoFeB|MgO heterostructures**. Transverse (a-d) and longitudinal (e-h) components of the current induced effective field, $\Delta H_T/J^{Under}$ and $\Delta H_L/J^{Under}$, respectively, are plotted as a function of the underlayer thickness for film stacks with different underlayers: (a,e) Ta, (b,f) TaN(Q=0.7%) (c,g) TaN(Q=1.2%) and (d,h) TaN(Q=2.5%). The effective field is normalized by the current density ($J^{Under}=1\times10^8$ A/cm$^2$) that flows into the underlayer. The solid and open symbols correspond to the effective field when the magnetization of the CoFeB layer is pointing along +Z and –Z, respectively.

**Figure S15. The offset field associated with the chiral magnetic texture: TaN|CoFeB|MgO heterostructures**. (a-d) The offset field $H_L^*$, i.e. the longitudinal field ($H_L$) at which the velocity becomes zero and the slope of domain wall velocity ($v_{DW}$) versus $H_L$ ($v_{DW}/H_L$), plotted as a function of underlayer thickness. The underlayer is (a,e) Ta, (b,f) TaN(Q=0.7%) (c,g) TaN(Q=1.2%) and (d,h) TaN(Q=2.5%). Solid and open symbols represent ↑↓ and ↓↑ domain walls, respectively. $H_L^*$ is evaluated when the wall is driven either by positive or negative voltage pulses: here, both results are shown together. The background color of each panel indicates the direction to which a corresponding domain wall moves; red: against the electron flow, blue: along the electron flow. (i) DMI parameter $D$ as a function of nitrogen gas

concentration ($Q$) when forming TaN using reactive sputtering of Ta.

**Table 1. Resistivity for different underlayers**. Film resistivity ($\rho$) of the underlayer in the heterostructure evaluated using the underlayer thickness ($d$) dependence of the resistance are shown. The atomic concentration of nitrogen in the TaN films is determined by Rutherford Backscattering Spectroscopy[1].

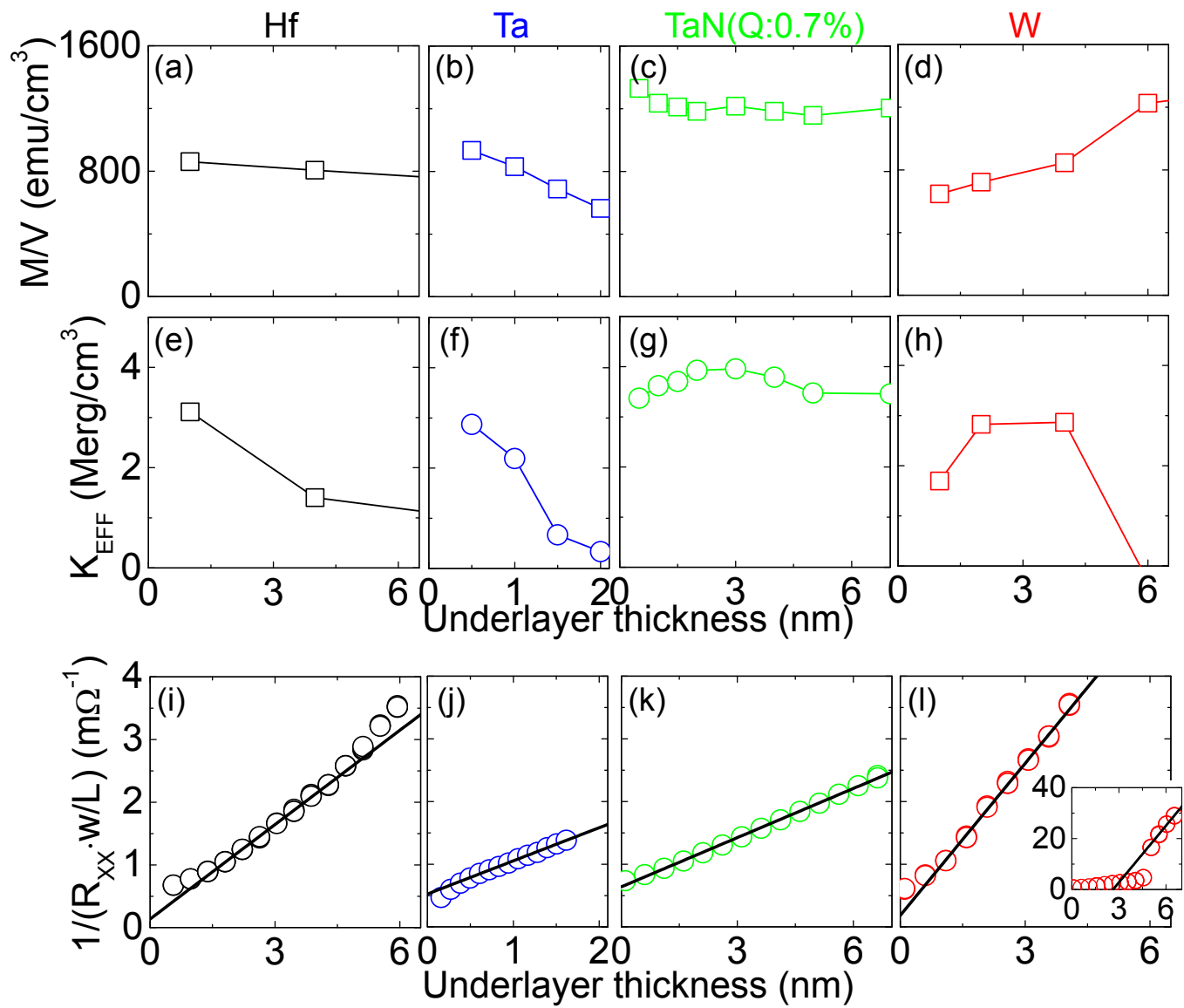

Fig. S1

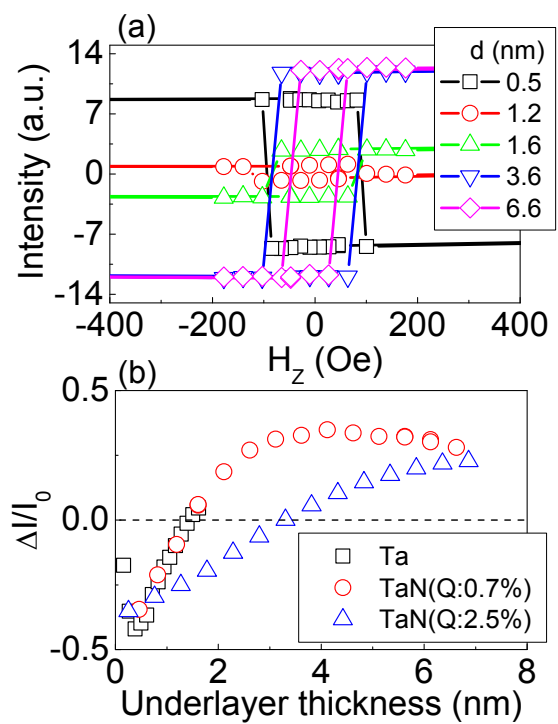

Fig. S2

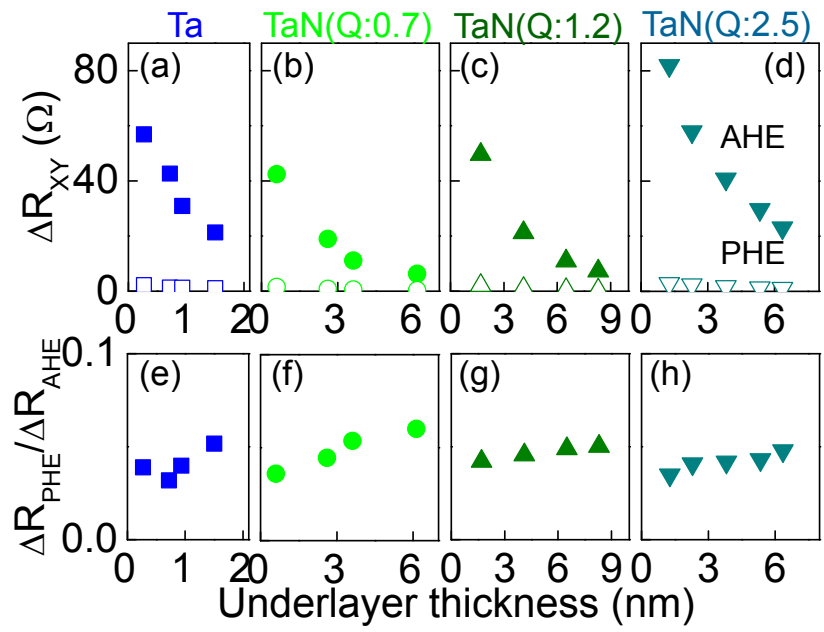

Fig. S3

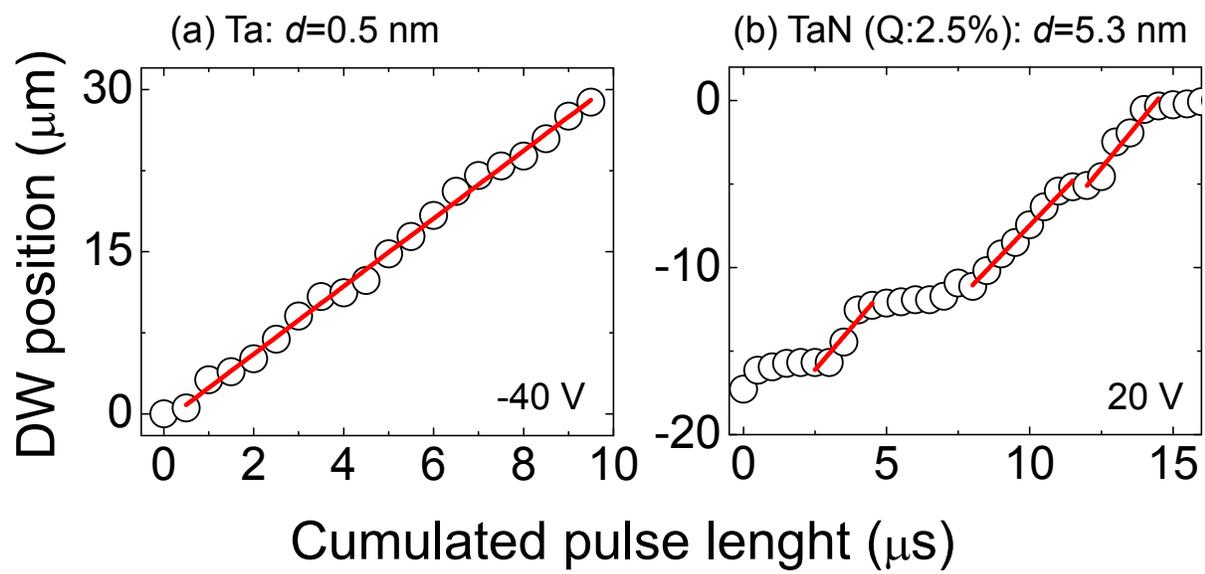

Fig. S4

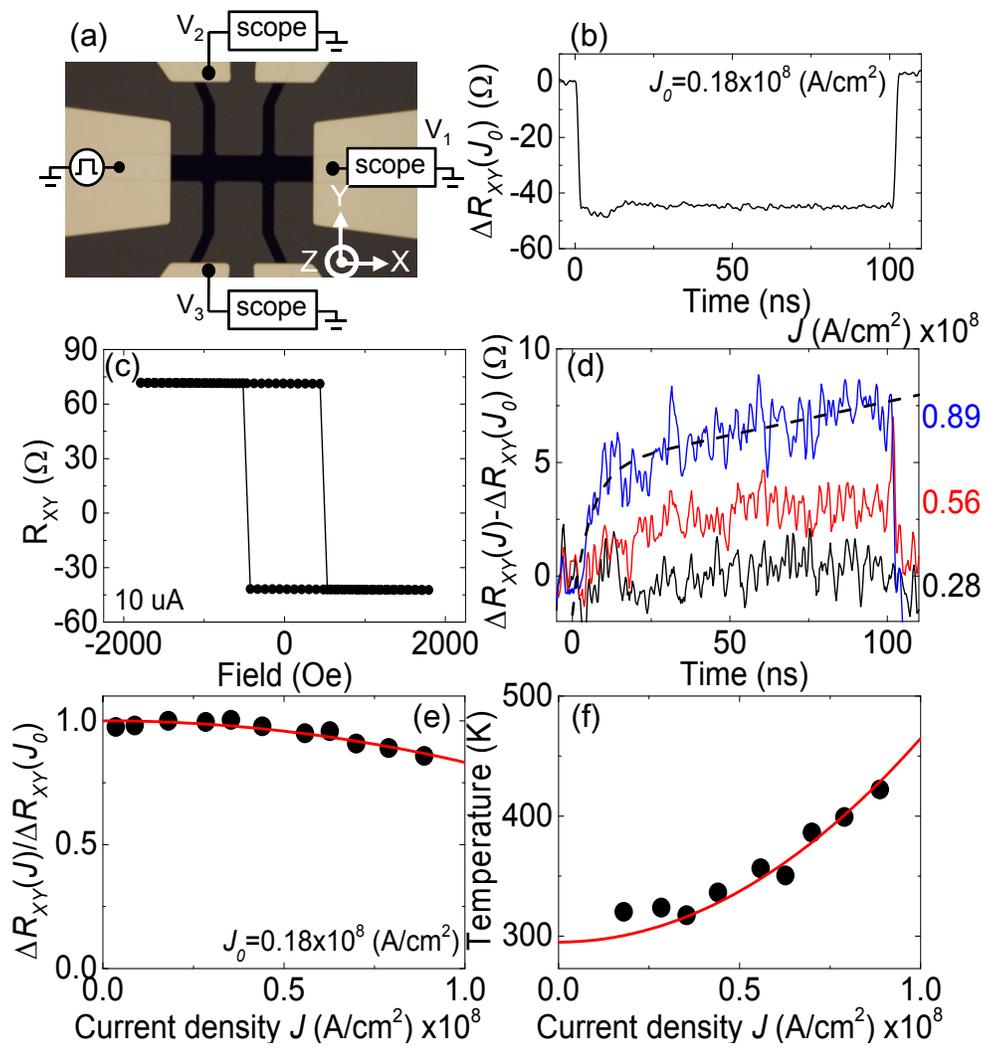

Fig. S5

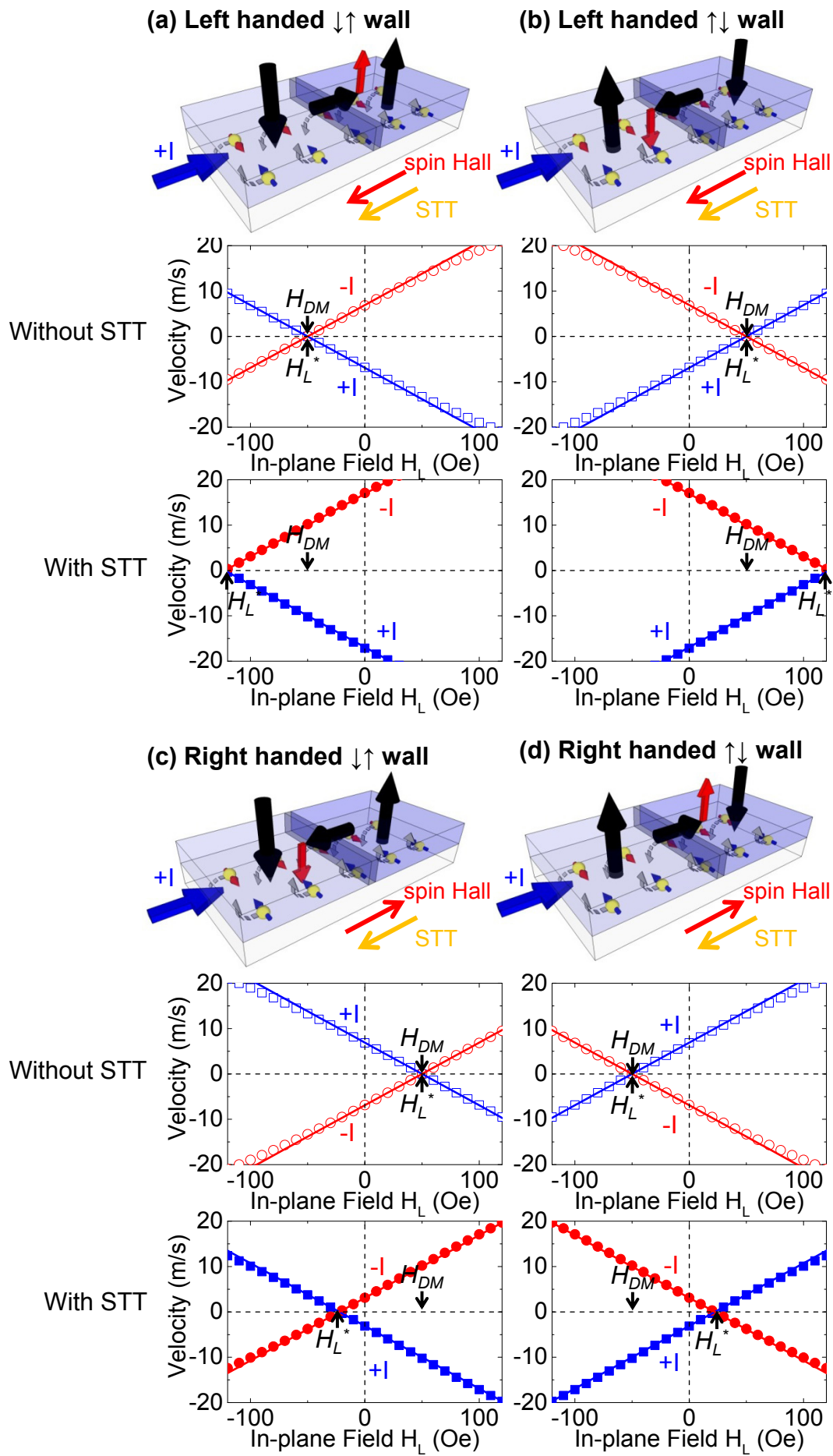

Fig. S6

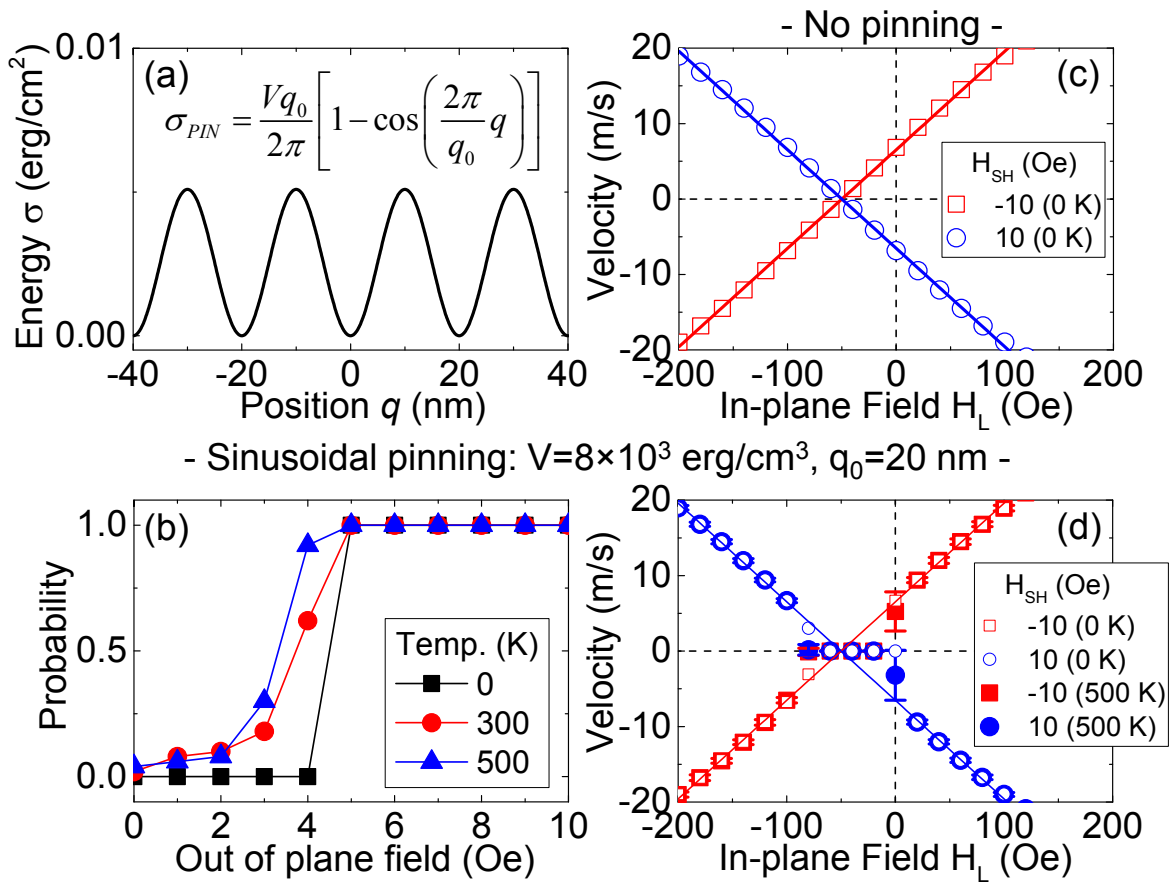

Fig. S7

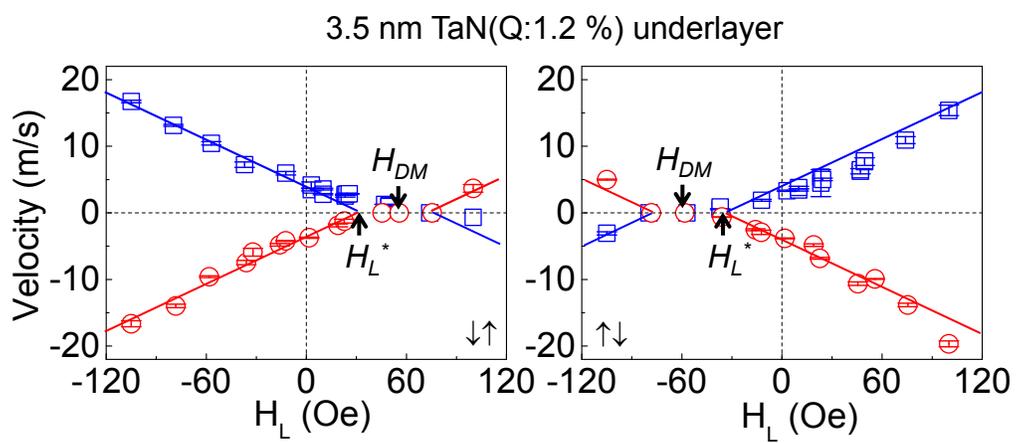

Fig. S8

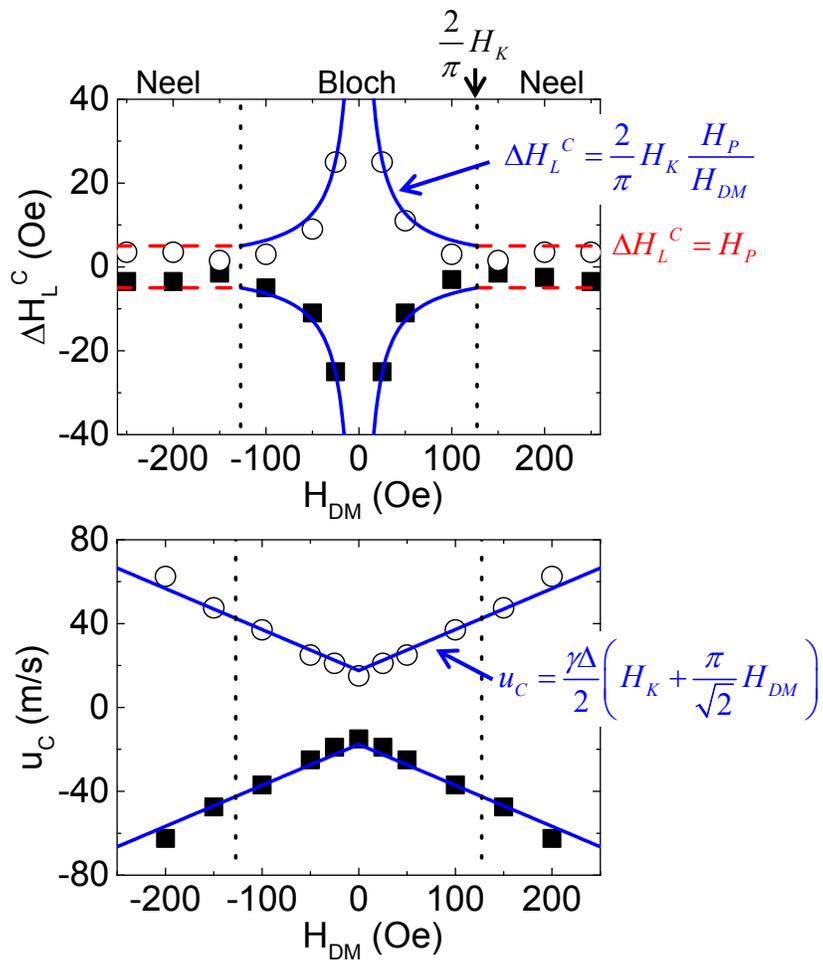

Fig. S9

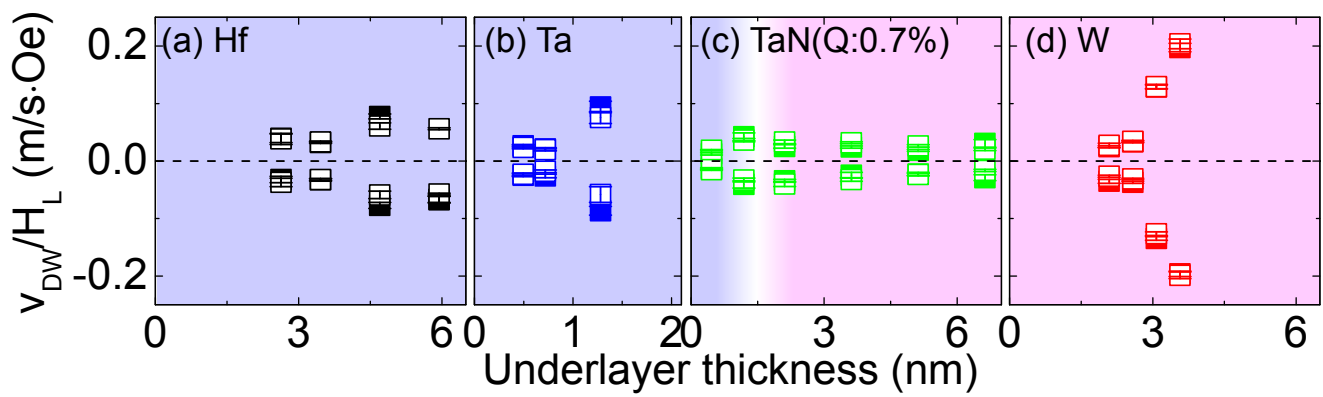

Fig. S10

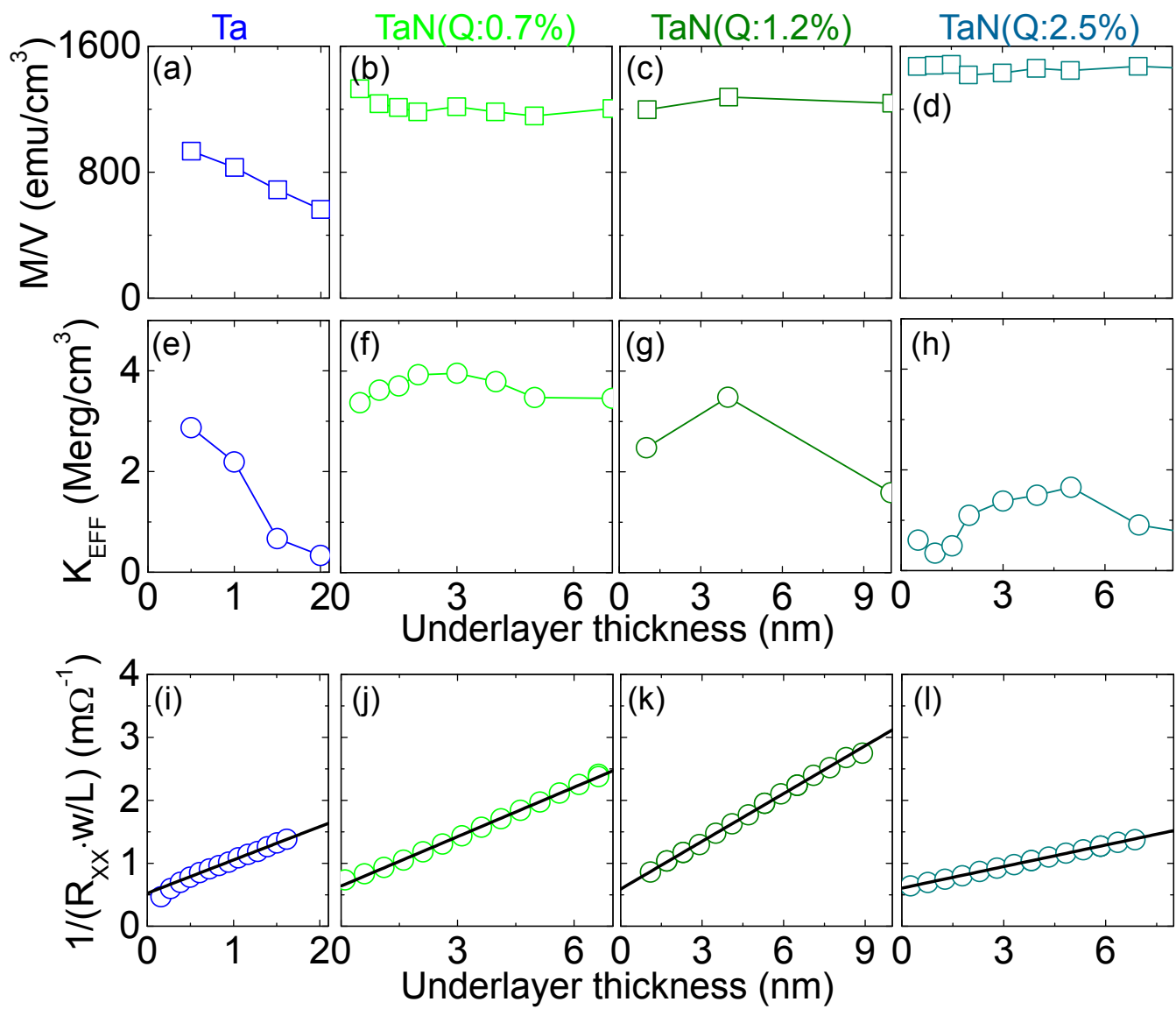

Fig. S11

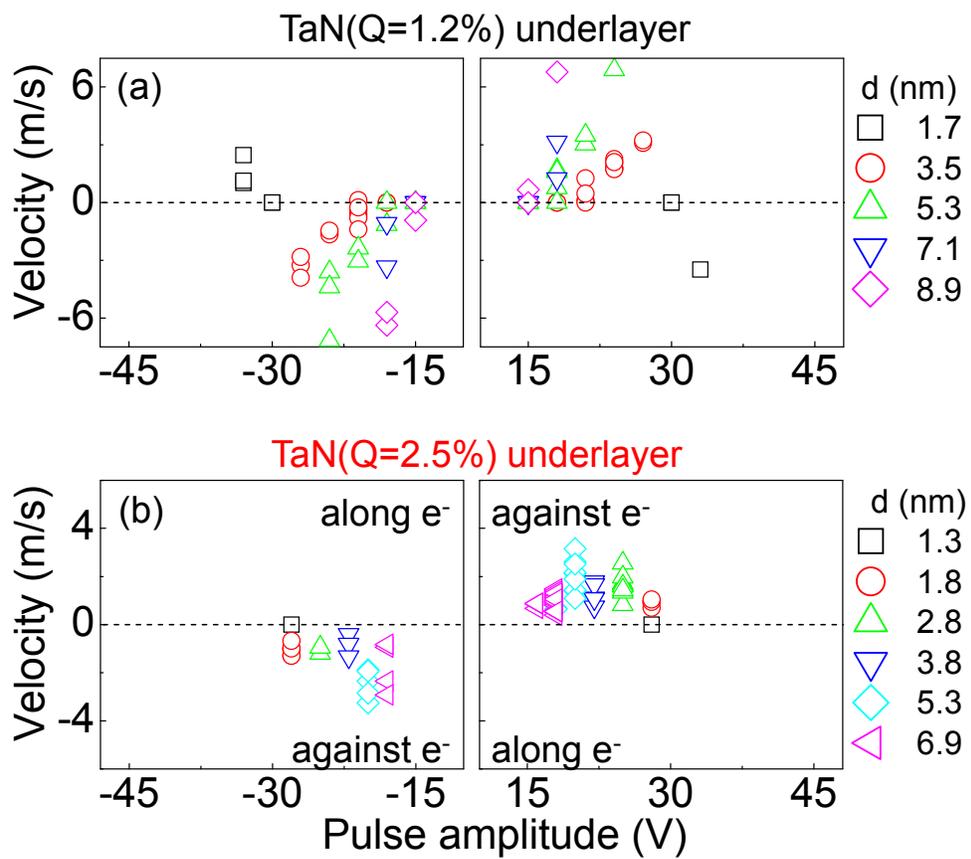

Fig. S12

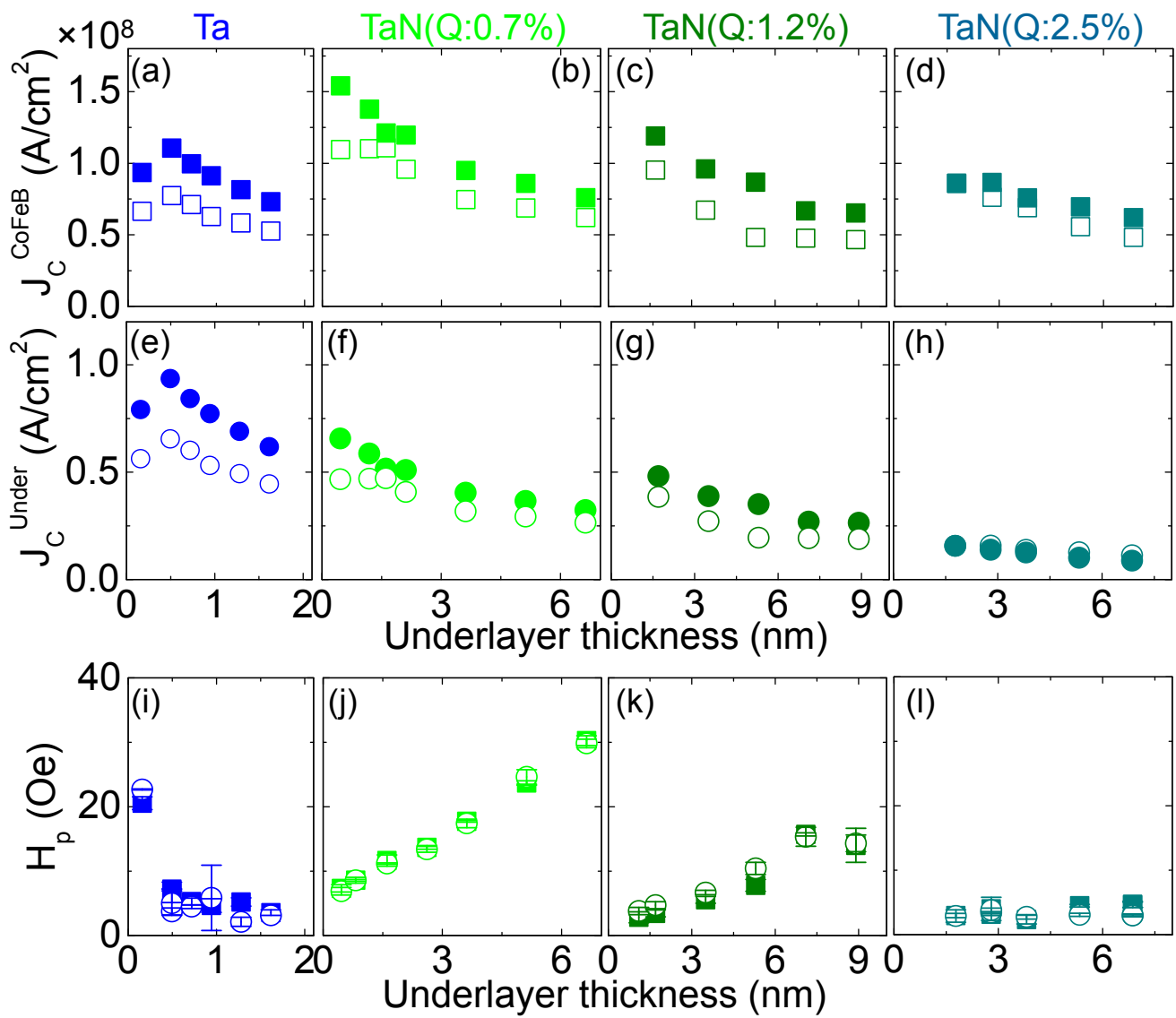

Fig. S13

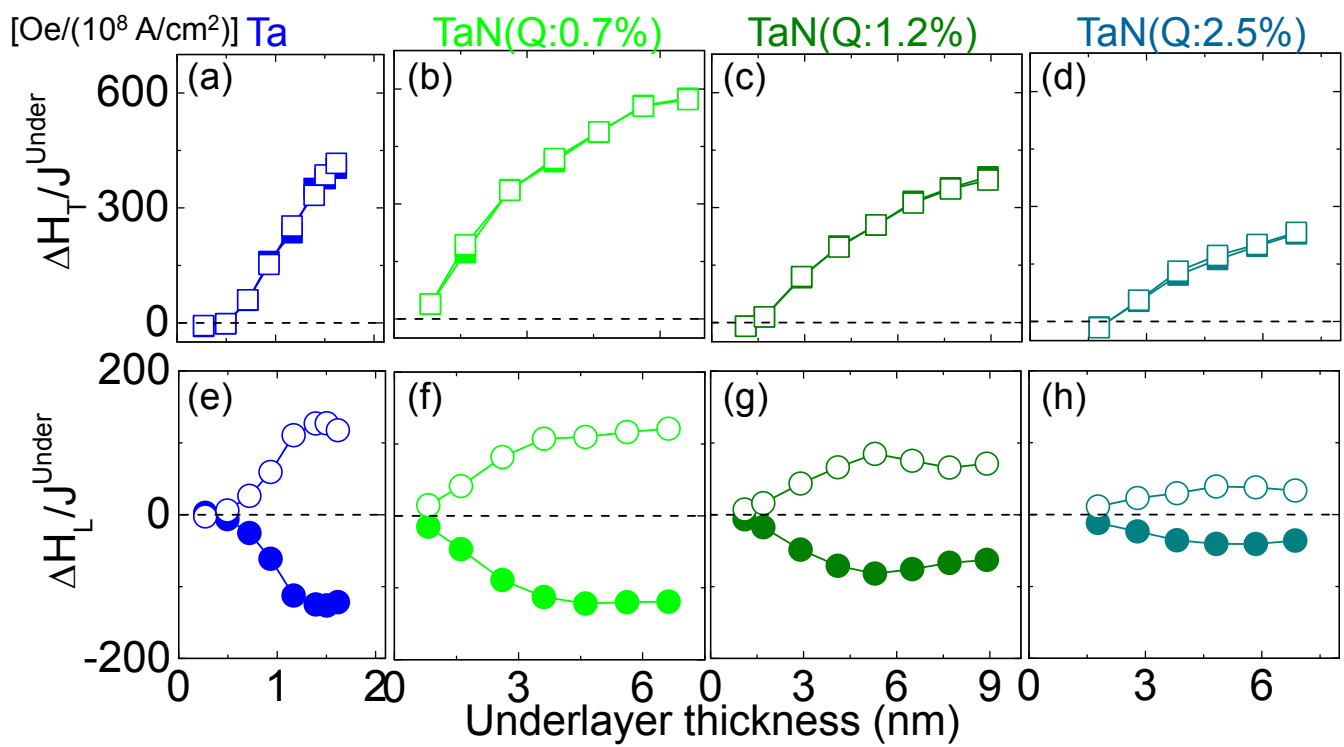

Fig. S14

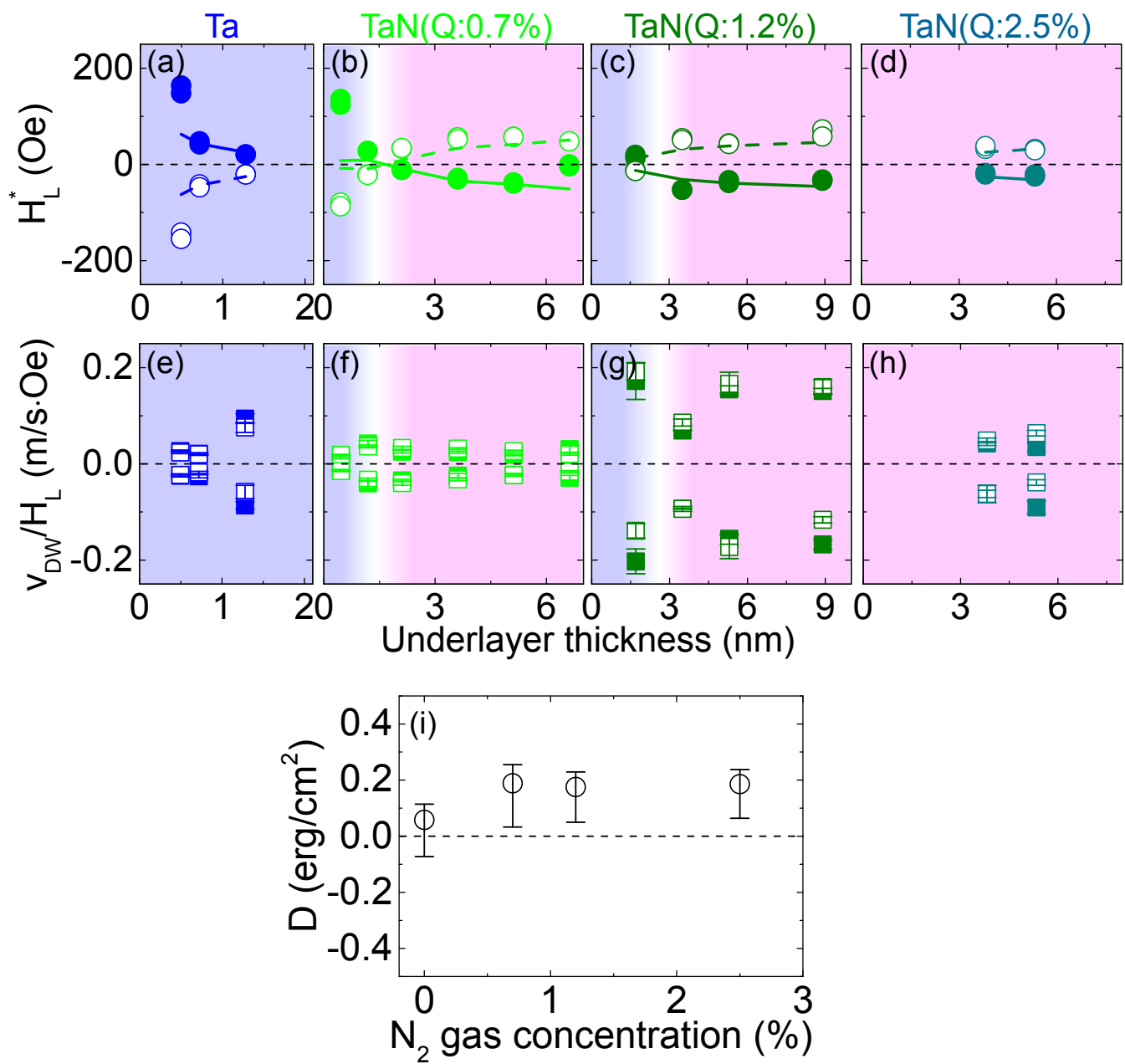

Fig. S15

|  | Hf | Ta | TaN(*Q*:0.7%) | TaN(*Q*:1.2%) | TaN(*Q*:2.5%) | W |
|---|---|---|---|---|---|---|
| **N at%** | N/A | 0 | 52±5 | 56±5 | 56±5 | N/A |
| **ρ (μΩ·cm)** | 199 | 189 | 375 | 395 | 876 | 124 |

Table 1